\pdfoutput=1
\documentclass[aps,prd,amsmath,floats,floatfix, twocolumn,
superscriptaddress,nofootinbib,showpacs,longbibliography]{revtex4-1}
\usepackage[T1]{fontenc}
\usepackage[utf8]{inputenc}
\usepackage{txfonts}
\usepackage{verbatim}
\usepackage[dvipsnames, usenames]{xcolor}
\definecolor{linkcolor}{rgb}{0.0,0.3,0.5}
\usepackage[hypertexnames=false, unicode, colorlinks=true, linkcolor=linkcolor,
citecolor=linkcolor, filecolor=linkcolor,urlcolor=linkcolor,
pdfusetitle]{hyperref}
\usepackage[all]{hypcap}
\usepackage{graphicx}
\usepackage{xspace}
\usepackage{amssymb}
\usepackage{microtype}
\usepackage{subfigure}

\usepackage{url}

\usepackage[english]{babel}
\usepackage{blindtext}

\usepackage[normalem]{ulem} 
\usepackage{bm} 
\usepackage{mathrsfs}
\usepackage{xspace} 
\usepackage{fontawesome}
\usepackage[normalem]{ulem}

\DeclareMathAlphabet{\mathpzc}{OT1}{pzc}{m}{it}

\usepackage[nomessages]{fp}

\newcommand{\sk}[1]{}

\newcommand{\vc}{\textcolor{linkcolor}{\texttt{Vishveshwara Catalog}}}

\begin{document}
\title{The shape of nonlinearity in binary-black-hole dynamics}

\author{Tousif Islam}
\email{tousif.islam@austin.utexas.edu}
\affiliation{Weinberg Institute, University of Texas at Austin, Austin, TX 78712, USA}

\hypersetup{pdfauthor={Islam}}

\date{\today}

\begin{abstract}
Black-hole perturbation theory remains remarkably accurate beyond its formal
small-mass-ratio regime, but what limits this accuracy remains
unclear. We develop a modular time-domain Teukolsky framework that isolates
finite-mass-ratio corrections to the orbital dynamics while keeping the
radiation at first perturbative order. Comparisons with numerical relativity show that the
dominant source of discrepancy changes across the waveform. For nonspinning
binaries, the available second-order dissipative flux removes $87\%$ of the
inspiral rate discrepancy, bringing it to within $\sim1\%$ of numerical
relativity. For precessing binaries, we identify a
$\nu^2\sin^2\iota$ dependence in the inspiral time rescaling that is not
explained by the leading orbit-averaged post-Newtonian terms,
suggesting finite-mass-ratio corrections to precessional dynamics. Near merger, the same
dissipative correction removes only $1$--$2\%$ of the discrepancy. We show
instead that the merger offset is quantitatively accounted for by the
difference between the initial Kerr background used in the perturbative
evolution and the remnant black hole, with $96\%$ of the leading predicted
frequency shift arising from the change in spin rather than mass. We release
$1571$ time-domain waveforms for $348$ binaries as the
\vc{}~\footnote{\href{https://tousifislam.com/VishveshwaraCatalog/}
{https://tousifislam.com/VishveshwaraCatalog/}}.
\end{abstract}

\maketitle

\noindent \textbf{\textit{Introduction: }} Black-hole perturbation theory
(BHPT) is formally organized around the small-mass-ratio limit
~\cite{Regge:1957td,Vishveshwara:1970cc,Zerilli:1970wzz,
Davis:1971gg,Teukolsky:1972my,Teukolsky:1973ha,Press:1973zz,Mino:1996nk,
Quinn:1996am,Poisson:2011nh,Barack:2018yvs,Pound:2021qin}, where the primary
black hole defines the background spacetime and the secondary is treated as
a point-particle perturbation. Yet this description remains remarkably
accurate well beyond its nominal regime
~\cite{vandeMeent:2020xgc,Islam:2023qyt,vandeMeent:2023ols,
Lousto:2010tb,Lousto:2010qx,Nakano:2011pb,Ramos-Buades:2022lgf}.
Post-adiabatic dynamics through second-order self-force reproduce
numerical-relativity (NR)
~\cite{Pretorius:2005gq,Campanelli:2005dd,Baker:2005vv}
inspirals for nonspinning binaries
~\cite{Pound:2019lzj,Wardell:2021fyy,Warburton:2021kwk,Miller:2020bft,
LeTiec:2011bk}, with extensions to spinning secondaries and slowly spinning
primaries showing similar promise
~\cite{Mathews:2021rod,Mathews:2025nyb,Mathews:2025txc,Akcay:2019bvk,
Piovano:2020zin}. In parallel, time/frequency, and amplitude mappings
carry perturbative waveforms toward comparable masses
~\cite{Rifat:2019ltp,Islam:2022laz,Rink:2024swg,Islam:2023qyt,
Islam:2023jak}.
What remains unclear is which finite-mass-ratio effects drive this agreement,
where they cease to be sufficient, and what ultimately limits the
perturbative description
~\cite{Islam:2025tjj,Islam:2023aec,Price:2022lcx}. We address these questions
by constructing the binary dynamics from independently switchable
perturbative ingredients and comparing the resulting waveforms with
NR. Identifying where the perturbative description succeeds and fails may
enable more accurate and efficient modeling at large mass ratios and
high spins, where NR simulations remain sparse but gravitational-wave (GW)
observations are beginning to probe the parameter space
~\cite{LIGOScientific:2021djp,LIGOScientific:2025obm,
LIGOScientific:2020zkf,Islam:2026oos}.

\begin{figure}
    \centering
    \includegraphics[width=\columnwidth]{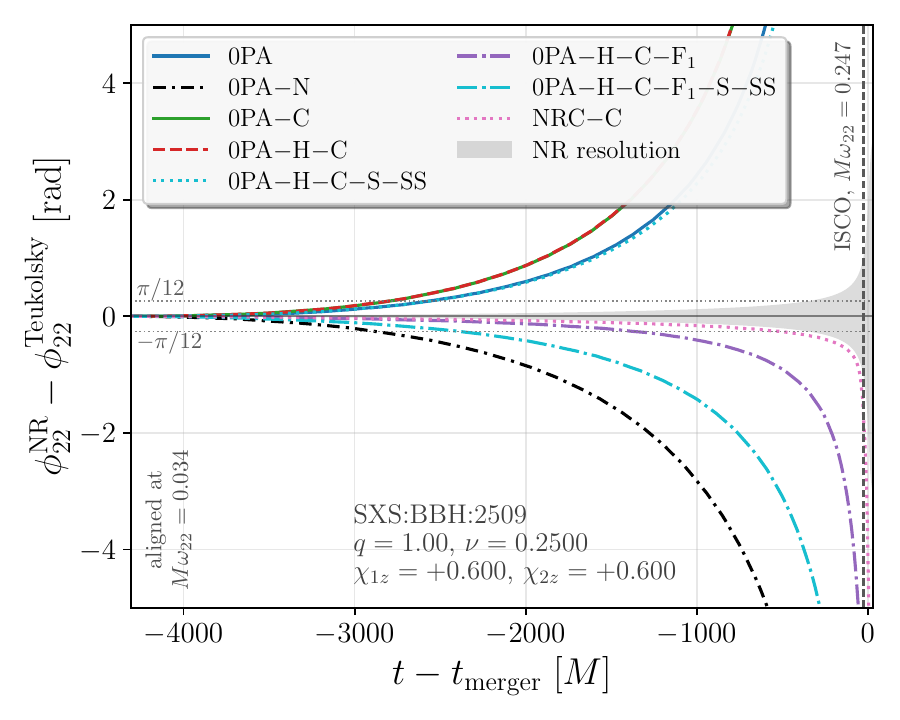}
    \caption{Building a BHPT waveform toward comparable masses. For the
    equal-mass, aligned-spin binary SXS:BBH:2509
    ($\chi_{1z}=\chi_{2z}=0.6$), we compare NR with a sequence of BHPT
    waveforms in which finite-mass-ratio corrections are added to the
    orbital dynamics through
    \textcolor{linkcolor}{\texttt{gwKerrDynamics}}, while the radiation is
    computed in every case with the same first-order
    \textcolor{linkcolor}{\texttt{gwTeukolsky}} evolution. The curves show
    the resulting $(2,2)$ phase difference from NR after alignment at
    $M\omega_{22}=0.034$. The improvement is not monotonic: conservative
    corrections initially increase the dephasing, while the second-order
    dissipative correction $F_1$ removes most of it. The gray band shows
    the NR uncertainty, and the vertical dotted line marks the
    Kerr ISCO.}
    \label{fig:ladder}
\end{figure}

\noindent \textbf{\textit{Perturbative framework: }}
We denote the primary and secondary masses by $m_1$ and $m_2$, with
$M=m_1+m_2$, $q=m_1/m_2\geq1$, and symmetric mass ratio
$\nu=m_1m_2/M^2$. At each instant, the secondary follows an osculating Kerr
geodesic~\cite{Gair:2010iv}, characterized by its specific energy $E$,
axial angular momentum $L_z$, and Carter constant $Q$~\cite{Carter:1968rr},
which we collect as $I^A=(E,L_z,Q)$. Away from orbital
resonances~\cite{Flanagan:2010cd}, their slow evolution and the orbital
phases admit the two-timescale expansion
~\cite{Hinderer:2008dm,Miller:2020bft}
\begin{equation}
\begin{aligned}
\dot I^A &= \nu F_0^A(I)+\nu^2F_1^A(I)+\mathcal{O}(\nu^3),\\[-2pt]
\dot\phi^\alpha &= \Omega^\alpha(I)+\nu\Omega_1^\alpha(I)
+\mathcal{O}(\nu^2).
\end{aligned}
\label{eq:forcing}
\end{equation}
Here $F_0^A$ and $F_1^A$ are the leading and next-order
radiation-reaction forces, while $\Omega^\alpha$ are the Kerr geodesic
frequencies~\cite{Schmidt:2002qk,Fujita:2009bp,Mino:2003yg} and
$\Omega_1^\alpha$ their first finite-mass-ratio corrections. Because
radiation reaction accumulates over $t\sim\nu^{-1}$, the phase has the
structure $\phi=\nu^{-1}\phi^{(0)}+\phi^{(1)}+\mathcal{O}(\nu)$
~\cite{Hinderer:2008dm,Pound:2021qin}. Thus $F_0^A$ determines the leading
accumulated phase, while the next $\mathcal{O}(1)$ phase depends on both
$F_1^A$ and $\Omega_1^\alpha$. This separation lets us introduce
finite-mass-ratio effects one at a time and determine which matter at
different stages of the binary evolution.

We compute the dissipative fluxes by solving the frequency-domain
Teukolsky equation for bound Kerr geodesics
~\cite{Teukolsky:1973ha,Teukolsky:1974yv,Sasaki:2003xr,Hughes:1999bq,
Drasco:2005kz}, defining our adiabatic baseline
\textcolor{linkcolor}{\texttt{0PA}}. We then add four classes of
finite-mass-ratio corrections independently. First, radiation absorbed by
the horizon~\cite{Teukolsky:1974yv,Poisson:1994yf,Hughes:2001jr} changes
the primary through $dM_b/dt=\dot E_{\rm H}$ and
$dJ_b/dt=\dot L_{\rm H}$, where $M_b$ and $J_b$ are its mass and angular
momentum, and $\dot E_{\rm H}$ and $\dot L_{\rm H}$ are the corresponding
horizon fluxes. Evolving the orbit on this changing Kerr background
defines \textcolor{linkcolor}{\texttt{H}}. Second, the conservative
correction \textcolor{linkcolor}{\texttt{C}} follows from the redshift
invariant and the first law of binary mechanics
~\cite{Detweiler:2008ft,LeTiec:2011ab,LeTiec:2015kgg,Fujita:2016igj,
Barack:2011ed,Shah:2012gu,Isoyama:2014mja,vandeMeent:2015lxa}. For
inclined and eccentric Kerr orbits, we compute the redshift by
metric reconstruction
~\cite{Chrzanowski:1975wv,Cohen:1974cm,Wald:1978vm,Pound:2013faa,
Merlin:2016boc,vanDeMeent:2017oet,vandeMeent:2017bcc,Nasipak:2025tby}.
Third, the next dissipative correction,
\textcolor{linkcolor}{\texttt{F1}}, comes from the second-order
Schwarzschild calculation
~\cite{Warburton:2021kwk,Pound:2019lzj,Wardell:2021fyy}, taken from the
Black Hole Perturbation Toolkit~\cite{BHPToolkit,WaSABI} and checked
against the known PN flux~\cite{Blanchet:2023bwj,Blanchet:2013haa}.
Since the Kerr result is not yet available, for spinning
primaries we evaluate this correction at matched orbital frequency and
test the consequences of this approximation below. For quasicircular
equatorial binaries, the distinction between these corrections is
transparent:
\begin{equation}
\frac{d\Omega}{dt}=\frac{dE/dt}{dE/d\Omega},
\qquad
\frac{d\phi}{dt}=\Omega .
\label{eq:freqdriven}
\end{equation}
Dissipative corrections change the flux $dE/dt$, while conservative
corrections change the energy--frequency relation $dE/d\Omega$. Finally,
when the secondary spins, we include its Mathisson--Papapetrou--Dixon
dynamics
~\cite{Mathisson:1937zz,Papapetrou:1951pa,Dixon:1970zza,Tulczyjew:1959,
Steinhoff:2009tk} in both the orbit and source
~\cite{Tanaka:1996ht,Saijo:1998mn,Harms:2015ixa,Akcay:2019bvk,
Mathews:2021rod,Drummond:2022xej,Skoupy:2023lih}, separating the
pole-dipole contribution linear in spin,
\textcolor{linkcolor}{\texttt{S}}, from the quadratic-in-spin contribution,
\textcolor{linkcolor}{\texttt{SS}}, which also includes the spin-induced
quadrupole of a Kerr secondary~\cite{Rahman:2026qho}. Each ingredient can
be switched on independently, allowing us to isolate its effect
on the orbital evolution and waveform.

We consider two extensions toward comparable masses that do not add a
new self-force order. First, we restore the Newtonian mass-asymmetry factor
of each radiative mode
~\cite{Thorne:1980ru,Kidder:2007rt,Blanchet:2013haa,Warburton:2021kwk,
Blanchet:2008je}, rescaling
$h_{\ell m}\rightarrow c_{\ell m}(\nu)h_{\ell m}$ and
$\mathcal{F}_{\ell m}\rightarrow
c_{\ell m}^2(\nu)\mathcal{F}_{\ell m}$,
where $c_{\ell m}$ is the function of $X_{1,2}=m_{1,2}/M$. Applying
the same factor to the orbital flux defines
\textcolor{linkcolor}{\texttt{0PA-N}}. Second, for nonprecessing binaries
we construct an all-$\nu$ flux from the exact $3.5$PN comparable-mass
series, including aligned-spin terms, and calibrate only the unknown
$4$PN-and-higher contribution to NR
~\cite{Blanchet:2013haa,Blanchet:2023bwj,Damour:1997ub},
\begin{equation}
\hat{\mathcal F}_{\rm NRC}
=\frac{P_3(v)}{Q_4(v)}
\left[1+x^4\sum_{k=0}^{3}
g_k(\nu,\chi_{\rm eff},\delta\chi_a)w^k\right],
\label{eq:nrcflux2}
\end{equation}
where $x=(M\Omega)^{2/3}$ and $v=\sqrt{x}$. When tested against SXS
nonprecessing simulations with
$1\leq q\leq20$~\cite{Boyle:2019kee,Scheel:2025jct}, the resulting flux
has an rms error of $0.94\%$. Its predicted
$\nu\rightarrow0$ limit agrees with our Kerr Teukolsky flux to $0.8\%$
outside $10M$. We denote this tier
\textcolor{linkcolor}{\texttt{NRC}}, or
\textcolor{linkcolor}{\texttt{NRC-C}} when \textcolor{linkcolor}{\texttt{C}}
is included.

We evolve each prescription through the adiabatic inspiral until the
two-timescale expansion breaks down near the innermost stable orbit
~\cite{Buonanno:2000ef,Ori:2000zn,Kesden:2011ma}. There the radial
potential becomes shallow and the separation between the orbital and
radiation-reaction timescales is lost, requiring a description
that connects the evolving inspiral to plunge. For aligned
quasicircular binaries, we use the composite transition of Comp\`ere and
K\"uchler
~\cite{Compere:2021zfj,Compere:2021iwh,Kuchler:2024esj}, which improves
the Ori--Thorne matching~\cite{Ori:2000zn,Compere:2019cqe} from
$\mathcal{O}(\nu^{3/5})$ to $\mathcal{O}(\nu)$ and connects smoothly to a
geodesic plunge~\cite{Dyson:2023fws}. For inclined quasicircular binaries,
we use its Mino-time generalization, matching at the innermost stable
spherical orbit~\cite{RibesMetidieri:2026tat,Apte:2019txp,Lim:2019xrb}
and constraining the energy through transition and plunge with the radial
potential. No transition solution is yet available for
eccentric motion, so there we connect the inspiral directly to a
constrained plunge~\cite{Gair:2010iv}. These constructions provide a
worldline $\{z^\mu(t),u^\mu(t)\}$ from inspiral through plunge,
which then becomes the source for the time-domain Teukolsky evolution.

The worldline provides the point-particle source for our
first-order time-domain Teukolsky evolution
~\cite{Teukolsky:1973ha,Sasaki:2003xr}. This means that
the orbital dynamics can include higher-order finite-mass-ratio effects,
while the radiative field remains linear in the perturbation.
We evolve the outgoing Newman--Penrose curvature scalar $\psi_4$, which
encodes the gravitational radiation, on the Kerr background using the
Kinnersley tetrad~\cite{Kinnersley:1969zza,Newman:1961qr}. We rescale the
field as $\Psi=\rho^{-4}\psi_4$, where $\rho$ is the
Kinnersley-tetrad spin coefficient, and use the axial symmetry of Kerr to
separate its azimuthal dependence,
$\Psi=\sum_m e^{im\tilde\phi}r^3\Phi_m(t,r_*,\theta)$. Here $r_*$ is the
Kerr tortoise coordinate and $\tilde\phi$ is a horizon-regular azimuthal
coordinate. This decomposition lets us evolve each $m$ mode separately,
turning the problem into a set of $2{+}1$-dimensional
time-domain evolutions
~\cite{Sundararajan:2007jg,Sundararajan:2008zm,Sundararajan:2010sr,
Burko:2016sfi}.
We evolve each $m$ mode with a three-stage strong-stability-preserving
Runge--Kutta scheme~\cite{Shu:1988,Gottlieb:2001}, using fifth-order radial
upwinding~\cite{Field:2020rjr} and a spectral representation in the polar
direction, with GPU acceleration
~\cite{McKennon:2012iq,Field:2020rjr}. In constructing the point-particle
source, we treat the azimuthal delta function analytically and reduce the
radial $\delta$, $\delta'$, and $\delta''$ terms before discretization, so
that the evolution never differentiates a distribution
~\cite{Sundararajan:2007jg}.
A hyperboloidal layer compactifies the exterior and carries
these evolutions directly to future null infinity $\mathscr I^+$
~\cite{Zenginoglu:2010cq,Zenginoglu:2011zz,Zenginoglu:2007jw,
Bernuzzi:2011aj,Vaswani:2026twx,Vishal:2023fye,Vishal:2025pqc,
Markakis:2023pfh,DaSilva:2023xif,PanossoMacedo:2022fdi,
PanossoMacedo:2024nkw}, avoiding an artificial outer boundary and
finite-radius waveform extrapolation. We read $\psi_4$ directly at
$\mathscr I^+$ and recover the strain by fixed-frequency integration
~\cite{Reisswig:2010di}. In our pipeline,
\textcolor{linkcolor}{\texttt{gwKerrDynamics}} constructs the
inspiral--transition--plunge worldline, which is passed to
\textcolor{linkcolor}{\texttt{gwTeukolsky}} to evolve the sourced field and
extract the waveform. Across tiers, only the orbital physics changes; the
source construction, Teukolsky evolution, and waveform extraction are held
fixed.

\begin{figure*}[t]
    \centering
    \includegraphics[width=\textwidth]{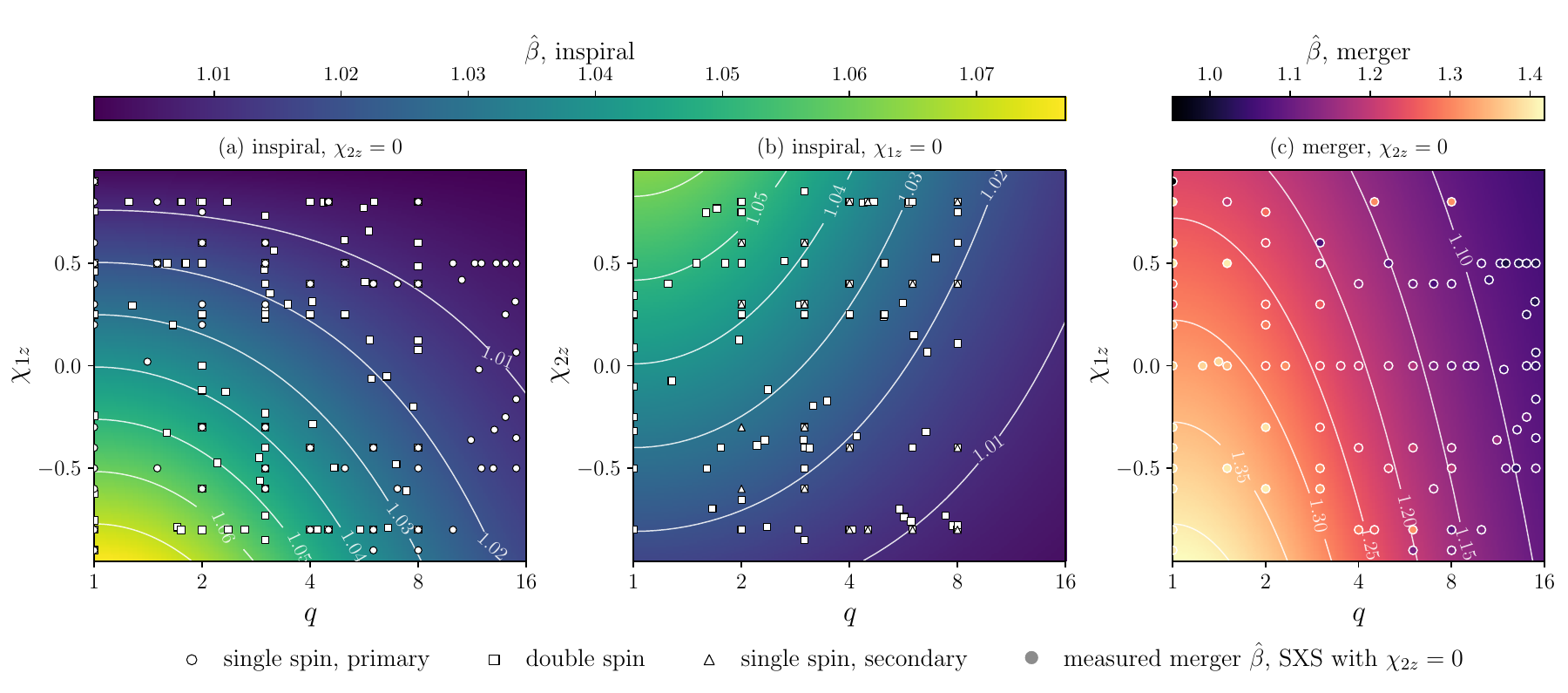}
    \caption{\label{fig:betamaps}
    Dependence of the fitted time rescaling $\hat\beta$ on mass ratio and
    spin. The left and middle panels show the inspiral rescaling when the
    secondary and primary spin are set to zero, respectively; the right
    panel shows the corresponding merger rescaling for $\chi_{2z}=0$.
    Symbols mark the configurations used in the analysis, with filled
    circles in the right panel showing the measured SXS merger values.
    The contours show the fitted relations. During inspiral, primary and
    secondary spins shift $\hat\beta$ in opposite directions, while the
    much larger merger rescaling is dominated by the mass and spin change
    of the remnant black hole.}
\end{figure*}

\noindent \textbf{\textit{Results: }} We apply this framework to $348$ nonspinning, spinning, and precessing
binaries spanning $1\leq q\leq100$ and spin magnitudes up to
$|\boldsymbol{\chi}_{1,2}|=0.95$. The parameter points are chosen to match
existing NR simulations across several catalogs. Across the dynamical tiers
described above, we generate $1571$ time-domain BHPT waveforms, evolved
with \textcolor{linkcolor}{\texttt{gwTeukolsky}} and publicly available
through the \vc{}.

Figure~\ref{fig:ladder} shows how these corrections compete in a
stringent case: SXS:BBH:2509, an equal-mass binary with
$\chi_{1z}=\chi_{2z}=0.6$, far from the small-mass-ratio limit. We align
the waveforms at $M\omega_{22}=0.034$, about $37$ GW cycles before merger,
and track
$\Delta\phi_{22}\equiv\phi_{22}^{\rm NR}-\phi_{22}^{\rm Teukolsky}$
as we change the dynamics driving the BHPT worldline. The result is
nonmonotonic. The \textcolor{linkcolor}{\texttt{0PA}} waveform
accumulates $6.3$ rad of dephasing by $500M$ before merger. Adding the
conservative correction increases this to $9.4$ rad, while horizon
absorption makes almost no difference. In contrast, the
second-order dissipative correction \textcolor{linkcolor}{\texttt{F1}}
reverses the sign and reduces the dephasing to $-0.9$ rad, with the
waveform remaining within $|\Delta\phi_{22}|<\pi/12$ until
$1360M$ before merger. This improvement comes from correcting
the orbital dynamics: the radiation is still generated by the same
first-order Teukolsky evolution. Thus, even at $q=1$, much of the inspiral
discrepancy can be removed without computing a second-order radiative
field. Secondary-spin effects further change this balance. Adding
\textcolor{linkcolor}{\texttt{S-SS}} without
\textcolor{linkcolor}{\texttt{F1}} improves the phase, but once
\textcolor{linkcolor}{\texttt{F1}} is included it instead increases the
dephasing from $-0.9$ to $-3.1$ rad at $-500M$. This is not
surprising: the available \textcolor{linkcolor}{\texttt{F1}} flux is
computed in Schwarzschild, whereas the primary here has spin
$\chi_{1z}=0.6$, outside its formal regime of validity. The NR-calibrated
\textcolor{linkcolor}{\texttt{NRC-C}} prescription gives the closest
agreement, remaining within $\pi/12$ until about $505M$ before merger, by
which point its phase difference is already comparable to the NR
uncertainty. As we show below, this competition is not unique to this
binary: during inspiral, the dissipative correction often provides the
largest improvement, but adding finite-mass-ratio effects one by one does
not necessarily move the BHPT waveform monotonically toward NR.

We next quantify the nonlinear information missing from the perturbative
waveforms. Previous comparisons
found that much of this difference can be captured by time and
amplitude rescalings
~\cite{Rifat:2019ltp,Islam:2022laz,Islam:2023qyt,Islam:2023jak,
Rink:2024swg,Islam:2025tjj}. Here we use the time rescaling as a diagnostic
of the missing dynamics, comparing our BHPT waveforms with their
SXS NR counterparts~\cite{Boyle:2019kee,Scheel:2025jct}. Since both
calculations use the same total mass $M$, we define
\begin{equation}
    t_{\rm NR}=\hat\beta\,t_{\rm Teuk},
    \qquad
    h_{\ell m}^{\rm NR}\simeq\alpha_{\ell m}h_{\ell m}^{\rm Teuk},
    \qquad
    \omega_{\rm NR}\simeq\omega_{\rm Teuk}/\hat\beta,
    \label{eq:calibration}
\end{equation}
where corresponding waveform quantities are evaluated at
$t_{\rm NR}=\hat\beta t_{\rm Teuk}$. The amplitude factor
$\alpha_{\ell m}$ accounts for the difference in waveform amplitude, while
$\hat\beta$ measures the evolution rate, with $\hat\beta=1$
corresponding to equal timescales. We measure $\hat\beta$ separately during
inspiral, transition, and merger. During inspiral, a fractional rate error
$\nu\Delta$ in $\dot x$, with $x=(M\Omega)^{2/3}$, gives
$\hat\beta-1=-(3/5)\nu\Delta+\mathcal{O}(\nu^2)$ from
Eq.~\eqref{eq:freqdriven}, so that $\hat\beta$ measures the
missing inspiral rate relative to NR.

We apply this diagnostic to the bare adiabatic dynamics,
\textcolor{linkcolor}{\texttt{0PA}}. For nonspinning binaries and those
with an aligned or tilted primary spin, where this tier contains the
leading-order Kerr flux, we find
\begin{equation}
    (\hat\beta-1)/\nu
    =A_0+A_1\chi_{1z}+A_2\nu+A_3\nu\sin^2\iota ,
    \label{eq:bare}
\end{equation}
where $\iota$ is the primary-spin tilt relative to the orbital angular
momentum. Fitting the aligned binaries gives
$A_0=0.618\pm0.016$, $A_1=-0.549\pm0.014$, and
$A_2=-0.94\pm0.09$.
The $A_2$ term is required at $10\sigma$, revealing a nonlinear
dependence on $\nu$ toward comparable masses, consistent with earlier
mappings of perturbative waveforms to NR~\cite{Rifat:2019ltp,Islam:2022laz}
but unlike conservative invariants, which extrapolate well under a
$\nu$ scaling~\cite{LeTiec:2011bk,Wardell:2021fyy}.
Keeping these coefficients fixed, the
tilted binaries give $A_3=3.74\pm0.33$, corresponding to
$\delta\hat\beta_{\rm tilt}\simeq A_3\nu^2\sin^2\iota$. This contribution
vanishes for aligned spins and grows toward comparable masses and larger
tilts.
Orbit-averaged PN theory does not explain this behavior. The leading
$2$PN tilt-dependent contribution
~\cite{Kidder:1995zr,Poisson:1997ha,Gergely:1999pd,Racine:2008qv},
which scales as $\chi_\perp^2$~\cite{Schmidt:2014iyl}, predicts an excess
of only $-0.006$, compared with the measured $+0.27$, and has the opposite
sign. Moreover, the data show no corresponding $\chi_\perp^2$ scaling.
The observed $\nu^2\sin^2\iota$ dependence instead points to a
finite-mass-ratio correction to precession about the total angular
momentum~\cite{Apostolatos:1994mx,Kidder:1995zr,Gerosa:2015tea}.

We include all finite-mass-ratio ingredients in our
framework except the second-order dissipative flux, using
\textcolor{linkcolor}{\texttt{0PA-H-C}} for nonspinning binaries and those
with only a spinning primary, and
\textcolor{linkcolor}{\texttt{0PA-H-C-S-SS}} whenever the secondary spins.
Across the nonprecessing binaries, the complicated
\textcolor{linkcolor}{\texttt{0PA}} dependence collapses to
\begin{equation}
    \hat\beta
    =1+\nu\left(a_0+a_1\chi_{1z}+a_2\chi_{2z}\right),
    \label{eq:fit}
\end{equation}
with $a_0=0.1591\pm0.0042$, $a_1=-0.1568\pm0.0080$, and
$a_2=0.0979\pm0.0088$, and an rms residual of $0.0127$ in $\hat\beta$.
Compared with the bare dynamics, the nonspinning coefficient falls from
$0.618$ to $0.159$, the magnitude of the primary-spin coefficient from
$0.549$ to $0.157$, and the nonlinear dependence on $\nu$ is no
longer required. The binaries for which we compute the full
conservative--dissipative ladder show where this reduction comes from:
the conservative correction \textcolor{linkcolor}{\texttt{C}} changes
$(\hat\beta-1)/\nu$ by $-0.333$ on average,
\textcolor{linkcolor}{\texttt{F1}} by $-0.177$, and horizon absorption
\textcolor{linkcolor}{\texttt{H}} by only $+0.003$. Together,
\textcolor{linkcolor}{\texttt{C}} and \textcolor{linkcolor}{\texttt{H}}
remove about two thirds of the bare discrepancy, lowering the mean
$(\hat\beta-1)/\nu$ from $0.50$ to $0.17$. Their relative importance
depends on the primary spin: the conservative correction is
twice as large as \textcolor{linkcolor}{\texttt{F1}} for nonspinning
binaries and five times as large for retrograde primaries, but only half
as large for prograde ones. It lowers $\hat\beta$ for nearly $90\%$ of
these binaries; the exceptions all have prograde primary spins with
$\chi_{1z}\geq0.5$, including SXS:BBH:2509 in Fig.~\ref{fig:ladder},
where the conservative correction increases the dephasing before
\textcolor{linkcolor}{\texttt{F1}} reverses it.

Figures~\ref{fig:betamaps}(a) and (b) show the relation in
Eq.~\eqref{eq:fit} across the primary- and secondary-spin planes. The
correction grows toward comparable masses, while aligned primary and
secondary spins act in opposite directions, decreasing and increasing
$\hat\beta-1$, respectively. The surface describes the nonspinning,
single-spin, and double-spin binaries. Once the known finite-mass-ratio
effects are included, the remaining discrepancy is predominantly an
$O(\nu)$ rate error with a spin dependence, as expected if the
dominant missing piece is the next-order dissipative forcing. This
relation is predictive: its coefficients can be determined from the
simpler spin families without using the double-spin binaries, with only a
$7\%$ increase in the overall residual when applied to them. Its form can
be understood from PN theory. For these tiers the conservative
energy--frequency relation is already correct through $O(\nu)$, so the
leading rate error comes from the missing $O(\nu)$ part of the normalized
flux, $\Delta=\left.\partial_\nu\ln\hat F\right|_{\nu\rightarrow0}$. At
$1$PN and $1.5$PN,
\begin{equation}
    \Delta=-\frac{35}{12}x+\frac{17}{4}\chi_1x^{3/2},
    \quad
    a_0^{\rm PN}=\frac{7}{4}\langle x\rangle,
    \quad
    a_1^{\rm PN}=-\frac{51}{20}\langle x^{3/2}\rangle .
    \label{eq:pncoefficients}
\end{equation}
Here the averages are taken over the same inspiral window used to measure
$\hat\beta$. Already at this order, PN reproduces $a_0$ to $8\%$, although
it accounts for only about half of the primary-spin coefficient. Including
the aligned-spin flux through $3.5$PN~\cite{LALSuite} largely closes this
gap: the $2.5$PN spin--orbit term brings $a_1^{\rm PN}$ to $94\%$ of the
measured value, while the $3.5$PN result reaches $110\%$. The nonspinning
coefficient remains $21$--$28\%$ higher than measured. Thus both the
mass-ratio and primary-spin dependence of $\hat\beta$ arise naturally from
finite-mass-ratio terms missing from the dissipative flux.

The second-order flux provides a direct test. Switching on
\textcolor{linkcolor}{\texttt{F1}} removes $87\%$ of
$|\hat\beta-1|$ for nonspinning binaries, $69\%$ when only the secondary
spins, and $43\%$ when both bodies spin. The improvement is smaller when
only the primary spins, where the Schwarzschild
\textcolor{linkcolor}{\texttt{F1}} flux removes just $19\%$. Measured
locally across the inspiral, the rate error is $3$--$7\%$
without \textcolor{linkcolor}{\texttt{F1}} and falls to within $1\%$
after it is included. Thus, for nonspinning binaries, the known
second-order dissipative flux accounts for most of the missing inspiral
rate. The weaker improvement for spinning primaries is
informative. The available \textcolor{linkcolor}{\texttt{F1}} flux is a
Schwarzschild result, which we evaluate at matched orbital frequency for
Kerr. It removes $43\%$ of the discrepancy for retrograde primaries but
increases it by $29\%$ for prograde primaries. The residual spin dependence
therefore points to physics absent from the Schwarzschild second-order
flux rather than to a generic failure of the perturbative waveform
calculation.

Finally, we ask whether $\hat\beta$ traces the shared orbital dynamics or
depends on the radiative mode. We repeat the measurement
for the $(2,1)$, $(3,3)$, $(3,2)$, $(4,4)$, $(4,3)$,
$(4,2)$, $(5,5)$, and $(5,4)$ modes. Their inferred rescalings differ
from that of the $(2,2)$ mode by at most $0.0017$, compared with a mean
$|\hat\beta-1|=0.025$. A single clock correction measured from the
$(2,2)$ mode therefore describes all of these modes, supporting the
interpretation that the dominant inspiral error lies in their shared
orbital dynamics rather than in the first-order radiative calculation.

Another way to diagnose the missing nonlinearity is to ask how the
residual rate difference scales with mass ratio, fitting
$|\hat\beta-1|\propto\nu^p$ with an exponent $p$ at different
stages of the evolution. For the nonspinning binaries, we find
$p=0.99$ during inspiral, consistent with a missing $O(\nu)$ correction to
the forcing, but $p=0.66$ through transition and $p=1.54$ near merger.
The change in scaling suggests that the dominant missing physics is
no longer the same. Here and below, the merger window extends from $100M$
before to $100M$ after the peak of $|h_{22}|$, or from the handover if it
occurs later. Across the nonprecessing binaries, the mean
$|\hat\beta-1|$ grows from $0.028$ during inspiral to $0.221$ near merger,
while \textcolor{linkcolor}{\texttt{F1}} removes only $1$--$2\%$ of the
merger offset, compared with $19$--$87\%$ during inspiral. The discrepancy
has therefore changed character. After handover, the Teukolsky field
continues to propagate on the initial Kerr background $(M,\chi_1)$,
whereas the NR waveform approaches the remnant $(M_f,\chi_f)$. Indeed, for
binaries with ringdown measurements, the model frequency differs
from the fundamental $(2,2,0)$ quasinormal frequency of its initial
background by a median of only $-0.0010$, with the majority agreeing
within $1\%$. The field is ringing at the frequency of the spacetime on
which it is evolved.

This interpretation also provides a way to understand the
mass-ratio and spin dependence of the merger discrepancy.
Across the nonprecessing binaries, the fitted merger time
rescaling follows
\begin{equation}
(\hat\beta_{\rm merger}-1)/\nu
=b_0+b_1\chi_{1z}+b_2\chi_{2z},
\label{eq:merger_law}
\end{equation}
with $(b_0,b_1,b_2)=(1.290,-0.402,0.195)$ and uncertainties
$(0.031,0.060,0.066)$, respectively
[Fig.~\ref{fig:betamaps}(c)].
Retrograde primaries show the largest offsets.
To understand the characteristic frequency difference
independently of the fitted time rescaling, we compare
the quasinormal frequencies of the initial and remnant
Kerr backgrounds. Defining $X_f=M_f/M$, the ISCO estimate
of Ref.~\cite{Buonanno:2007sv} gives $X_f=1-\nu\tilde E$ and 
\begin{equation}
\chi_fX_f^2=\chi_1X_1^2+\nu\tilde L,\quad
R_{\rm QNM}\equiv
\frac{\omega_{\rm NR}}{\omega_{\rm Teuk}}
\simeq\frac{\omega_{220}(\chi_f)}
{\omega_{220}(\chi_1)X_f}.
\label{eq:remnant}
\end{equation}
Since $X_1^2=1-2\nu+O(\nu^2)$, defining
$\delta\chi=\tilde L-2\chi_1(1-\tilde E)$ gives
$\chi_f=\chi_1+\nu\delta\chi+O(\nu^2)$ and hence
\begin{equation}
R_{\rm QNM}-1
=\nu\left[\tilde E+\delta\chi\,
\partial_\chi\ln\omega_{220}\right]_{\chi_1}
+O(\nu^2).
\label{eq:merger_smallnu}
\end{equation}
For Schwarzschild this predicts $1.46\nu$ for the
fractional frequency shift, with $96\%$ arising from
the change in spin rather than mass. Across the catalog,
the predicted coefficient is $1.48-0.46\chi_{1z}$,
comparable in magnitude and spin dependence to the
measured merger-rescaling coefficients.
The term $-2\chi_1(1-\tilde E)$ is essential:
without the reduction of the primary's share of $M^2$,
the predicted spin trend has the wrong sign.
Secondary spin enters this estimate only at higher order.
For precessing binaries, the same relation with
$\chi_{1z}=|\chi_1|\cos\iota$ describes the mean
behavior, although with larger scatter.
The ringdown provides a direct test of this picture,
independent of the fitted $\hat\beta_{\rm merger}$.
For binaries with reliable NR ringdown frequencies,
the median measured shift is
$\omega_{\rm NR}/\omega_{\rm Teuk}-1=0.209$,
compared with the parameter-free prediction $0.238$
from Eq.~\eqref{eq:remnant}.
A regression of the measured against predicted shifts
gives a slope of $0.80$ and an intercept of $0.02$,
with the majority agreeing within $10\%$.
Thus the large merger discrepancy does not mean that
the first-order Teukolsky field has ceased to ring
correctly. It rings consistently with the initial
Kerr background; the dominant missing effect is
that the binary has produced a different black hole.

Overall, the departure from the perturbative limit has a distinct physical
structure: finite-mass-ratio forcing controls the inspiral, precessional
dynamics introduces new nonlinear dependence, and the difference between
the initial and remnant Kerr backgrounds controls the merger.

\noindent \textbf{\textit{Acknowledgments: }}
We thank Digvijay Wadekar, Gaurav Khanna, Scott Field, Aaron Zimmerman,
Deirdre Shoemaker, Ajit Kumar Mehta, Tejaswi Venumadhav, Mark Cheung,
Matias Zaldarriaga, Barak Zackay, and Javier Roulet for helpful discussions and comments.
Computational resources were provided by Purdue Anvil through ACCESS award
PHY260280, the NAIRR Pilot through award NAIRR260298, and Stampede3 at the
Texas Advanced Computing Center (TACC) through allocation AST26023.
We acknowledge the use of GPT-5.6 Sol and Claude Opus 4.8 and 5.5 in
conversational and agentic workflows for software development, code
execution, and grammatical editing. All scientific results
were independently verified by the author and validated against established
results and data products in the field.

\appendix

\section{Overview of the appendices}
The appendices provide further details on the calculations presented in
the main text. Appendix~\ref{app:catalog} describes the \vc{} and its
coverage, while Appendix~\ref{app:closedforms} collects the closed-form
perturbative inputs and comparable-mass extensions. We present external
validation of \textcolor{linkcolor}{\texttt{gwKerrDynamics}} and
\textcolor{linkcolor}{\texttt{gwTeukolsky}} in
Appendix~\ref{app:validation}, followed by numerical resolution studies of nonspinning and
eccentric, inclined coalescences in Appendix~\ref{app:resolution}.
Appendix~\ref{app:ladder-q15} gives a second example of the perturbative
ladder at larger mass ratio. Finally, Appendix~\ref{app:agentic} describes
how the two codes were developed with AI agents and the validation strategy
used throughout that process.

\section{The Vishveshwara Catalog}
\label{app:catalog}

The \vc{}\footnote{\href{https://tousifislam.com/VishveshwaraCatalog/}
{https://tousifislam.com/VishveshwaraCatalog/}} is organized around a
tiered description of binary-black-hole (BBH) dynamics. Finite-mass-ratio effects,
including horizon absorption, conservative corrections, second-order
dissipation, and secondary-spin effects, can be added individually while
the radiative calculation is kept fixed. The orbital dynamics are generated
with \textcolor{linkcolor}{\texttt{gwKerrDynamics}}, and the corresponding
time-domain Teukolsky waveforms with
\textcolor{linkcolor}{\texttt{gwTeukolsky}}. This structure allows the
effect of individual perturbative ingredients to be isolated for the same
physical binary. The catalog is named after C.~V.~Vishveshwara, whose work
established the stability of the Schwarzschild black hole and helped lay the
foundations of black-hole ringdown~\cite{Vishveshwara:1970cc}.
We describe the agentic development, validation, and progress of
\textcolor{linkcolor}{\texttt{gwKerrDynamics}} and
\textcolor{linkcolor}{\texttt{gwTeukolsky}} in
Appendix~\ref{app:agentic}, with detailed external code validations presented in
Appendix~\ref{app:validation}.

\begin{table*}[t]
\caption{\label{tab:catalog}
Content of the \vc{}. Durations and GW cycle counts are medians, with the
full range in brackets, measured to the peak of $|h_{22}|$. Every waveform
continues for approximately $330M$ beyond the peak. The four full-ladder
configurations are subsets of the physical families and are listed separately
because they contain additional dynamical tiers. More details are in Appendix~\ref{app:catalog}.}
\begin{ruledtabular}
\begin{tabular}{lrrlll}
Family & Binaries & Waveforms & Parameter range & Modes &
Time [$M$] / GW cycles \\
\hline
Nonspinning
& 29 & 201 & $q=1$--$100$
& 29 & $9400\,[3600,32800]$ / $70\,[29,174]$ \\

Spinning primary
& 72 & 439 & $q=1$--$32$, $|\chi_{1z}|\leq0.9$
& 29 & $8900\,[2800,25100]$ / $63\,[29,156]$ \\

Spinning secondary
& 39 & 234 & $q=2$--$15.1$, $|\chi_{2z}|\leq0.81$
& 29 & $6100\,[4700,11500]$ / $49\,[38,81]$ \\

Double spin
& 120 & 491 & $q=1$--$8.6$, $|\chi_{1z,2z}|\leq0.95$
& 29 & $7300\,[3500,18000]$ / $53\,[31,112]$ \\

Precessing
& 76 & 148
& $q=1$--$15$, $|\boldsymbol{\chi}_1|\leq0.9$,
  $\iota=19^\circ$--$150^\circ$
& 12
& \begin{tabular}{l}
  $\nu$: $13900\,[3700,66400]$ / $84\,[30,214]$ \\
  $\epsilon$: $9400\,[2100,29900]$ / $60\,[21,170]$
  \end{tabular} \\

Full ladder
& 4 & 50 & $q=1$--$15$, $|\chi_{1z,2z}|\leq0.8$
& 29 & --- \\

Demonstration
& 8 & 8 & $q=12.1$, $e_0\leq0.2$, $\cos\iota_0\geq0.4$
& --- & --- \\
\hline
Total & 348 & 1571 & & &
\end{tabular}
\end{ruledtabular}
\end{table*}

The current release contains $348$ binary configurations and $1571$
time-domain Teukolsky waveforms, spanning $1\leq q\leq100$, spin magnitudes
up to $0.95$, and nonspinning, aligned-spin, precessing, and eccentric
systems (Table~\ref{tab:catalog}). Each physical configuration is assigned a
unique \textcolor{linkcolor}{\texttt{TEUK:BBH:NNNNN}} identifier, while
different dynamical tiers are stored as separate waveforms of the same
binary. This organization allows individual perturbative ingredients to be
compared without changing the underlying binary parameters.

Most configurations are chosen to match existing NR simulations: $301$ from
SXS~\cite{Boyle:2019kee,Scheel:2025jct}, $22$ from
MAYA~\cite{Ferguson:2023vta}, and $11$ from RIT~\cite{Healy:2022wdn}.
The RIT configurations extend the NR-matched nonspinning sample to $q=16$
and the spinning-primary sample to $q=32$. Two further configurations match
private NR simulations from \textsc{bam} at
$q=18$~\cite{Bruegmann:2006ulg} and Dendro-GR at
$q=24$~\cite{Fernando:2022php}. For these systems, the binary parameters are
taken directly from the corresponding NR simulations. Three additional
nonspinning configurations at $q=50$, $75$, and $100$, together with seven
$q=12.1$ demonstration systems, extend the catalog beyond the NR-matched
sample.

\begin{figure}[t]
    \centering
    \includegraphics[width=\columnwidth]{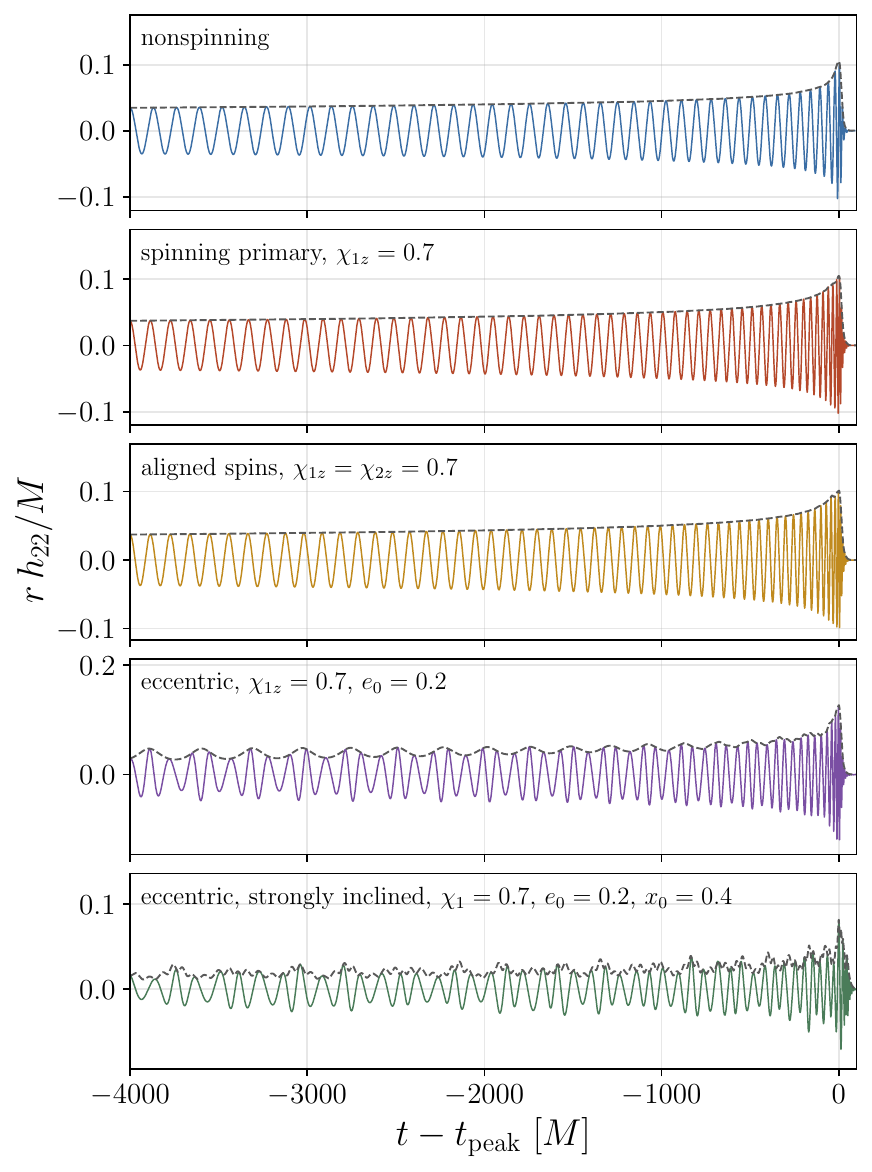}
    \caption{\label{fig:catalog-demo}
    Examples of $(2,2)$ strain waveforms in the \vc{}. Alongside
    nonspinning, spinning-primary, and aligned double-spin binaries, we
    show two $q=12.1$ demonstrations extending to eccentric and strongly
    inclined orbits. All waveforms are shifted so that the peak of
    $|h_{22}|$ occurs at $t=0$. More details are provided in Appendix~\ref{app:catalog}.}
\end{figure}

For aligned binaries, we independently evolve $29$ modes with
$2\leq\ell\leq9$ and $1\leq m\leq\ell$; negative-$m$ modes follow from
symmetry. For precessing binaries, all $m$ modes, including $m=0$, are
evolved for $2\leq\ell\leq3$, giving $12$ modes and allowing rotation to a
co-precessing frame~\cite{OShaughnessy:2011pmr,Boyle:2013nka}. Both
$\psi_4^{\ell m}$ and the strain $rh^{\ell m}/M$ are provided at future
null infinity. Quantities are expressed in units of the total mass $M$,
except for the \textcolor{linkcolor}{\texttt{0PA\_eps}}
waveforms. These use the primary mass $m_1$ and the small-mass-ratio
parameter $\epsilon=m_2/m_1$, rather than the symmetric-mass-ratio
parametrization used for the corresponding
\textcolor{linkcolor}{\texttt{0PA}} waveforms. 
The two prescriptions allow the effect of this choice of mass-ratio
parametrization to be assessed within the same leading-order dynamics.
Representative waveforms spanning several of the catalog families are
shown in Fig.~\ref{fig:catalog-demo}.

The production campaign used TACC Stampede3 and Lonestar6, Purdue Anvil,
UMass Unity, and NCSA Delta. Generating the $1571$ waveforms required
approximately $2\times10^5$ CPU core-hours. Precomputed flux and redshift
tables required a further $\approx2\times10^4$ CPU core-hours, while
resolution studies used $\approx3.5\times10^4$ CPU core-hours and
$\approx140$ GPU-hours. GPU acceleration on Anvil was also used for the
\textcolor{linkcolor}{\texttt{0PA\_eps}} precessing waveforms. Each catalog
entry records the binary parameters, corresponding NR simulation where
available, dynamical tier, available modes, and file information. The
waveform files contain the time series, available $\psi_4^{\ell m}$ and
$h^{\ell m}$ modes, and the metadata needed to identify the physical and
numerical configuration.

The present release is intended as a starting point for broader coverage of
the BBH parameter space. Future extensions will target more
extreme spins and larger mass ratios, where NR simulations become
increasingly expensive and perturbative calculations are particularly
valuable. Such coverage can support waveform models that combine information
from NR and BHPT, as well as template banks for intermediate-mass-ratio
binaries and other searches in regions where dense NR coverage is difficult
to obtain. Scripts to download and load waveforms from the \vc{} are provided through the publicly available \textcolor{linkcolor}{\texttt{gwModels}} package~\footnote{\href{https://github.com/tousifislam/gwModels}{https://github.com/tousifislam/gwModels}}.

\section{Closed-form fluxes and finite-mass-ratio prefactors}
\label{app:closedforms}

For the nonspinning perturbative inputs, we construct compact analytic
representations of the point-particle flux~\cite{Detweiler:1977gy,Poisson:1993vp,Tagoshi:1993dm,Cutler:1994pb,Sasaki:2003xr}, the first-order redshift
correction~\cite{Detweiler:2008ft,Akcay:2012ea}, and the available second-order non-spinning dissipative flux~\cite{Pound:2019lzj,Warburton:2021kwk}. We use $x=(M\varpi)^{2/3}$, with $\varpi=\dot\Phi_{22}/2$, and define
$s=\sqrt{1-3x}$ and
$\hat F=\mathcal F/[(32/5)\nu^2x^5]$, so that
$\hat F=\hat F_0+\nu\hat F_1+O(\nu^2)$. We write
\begin{align}
\hat F_0 &=
\frac{1+(s-1)\sum_{k=0}^{5}a_k s^k}{1-3x},
\nonumber\\
z_1 &=
\frac{2x\left[1+(s-1)\sum_{k=0}^{5}c_k s^k\right]}{1-3x},
\nonumber\\
\hat F_1 &=
\frac{x}{1-3x}
\left[-\frac{35}{12}
+(s-1)\sum_{k=0}^{5}p_k u^k\right],
\label{eq:closed_inputs}
\end{align}
where $u=2(s-s_-)/(s_+-s_-)-1$, with $s_+=0.9695$ and
$s_-=0.7211$, maps the interval $6.25M\leq r\leq50M$ covered by the
second-order tables onto $[-1,1]$.~\cite{Warburton:2021kwk} The variable
$s=\sqrt{1-3x}=1/u^t$ makes the light-ring behavior explicit~\cite{Akcay:2012ea}, while the
form chosen for $\hat F_1$ preserves the exact $1$PN coefficient
$-35/12$~\cite{Blanchet:2013haa}. These expressions reproduce the underlying data to
$3.3\times10^{-3}$ for $\hat F_0$ (24 radii spanning
$3.25M\leq r\leq50M$), $1.2\times10^{-3}$ for $z_1$ (91 nodes spanning
$3.03M\leq r\leq150M$), and $3\times10^{-2}$ for $\hat F_1$ over its
available range $r\geq6.25M$. The production waveforms use direct
interpolation of the underlying perturbative tables; the expressions above
provide closed-form approximations to the same inputs.

For the \textcolor{linkcolor}{\texttt{NRC}} tier, we construct a
closed-form flux that extends the dynamics toward comparable masses. We
start from a bounded Pad\'e resummation~\cite{Damour:1997ub} of the known spin-dependent PN flux through $3.5$PN~\cite{Blanchet:2013haa,Bohe:2013cla,LALSuite} and use available NR data to constrain additional terms beginning at $4$PN,
\begin{equation}
\hat F_{\rm NRC}
=
\frac{P_3(v)}{Q_4(v)}
\left[
1+x^4\sum_{k=0}^{3}g_k
\left(\frac{x-0.19}{0.13}\right)^k
\right],
\qquad v=\sqrt{x}.
\label{eq:nrc_flux}
\end{equation}
The coefficients $g_k$ depend on $\nu$, $\chi_{\rm eff}$, and the
antisymmetric spin combination and are calibrated to $151$ NR simulations
spanning $1\leq q\leq30$ and
$-0.95\leq\chi_{\rm eff}\leq0.98$. The resulting flux has an rms error of
$0.94\%$, increasing to $1.02\%$ when each simulation is excluded from its
own fit. Because the NR-calibrated terms enter only at $4$PN and beyond,
all known PN terms are preserved.

\begin{table*}[t]
\caption{\label{tab:validation}
External validation of the waveform pipeline. We test the orbital dynamics,
perturbative fluxes, conservative and secondary-spin corrections,
time-domain Teukolsky evolution, inspiral dynamics, and the complete
waveform against analytic results, published calculations, or independent
codes. The final column gives the largest relative difference over the
cases tested unless otherwise stated. More details are provided in Appendix~\ref{app:validation}.}
\footnotesize
\begin{ruledtabular}
\begin{tabular}{p{0.18\textwidth}p{0.34\textwidth}p{0.27\textwidth}p{0.14\textwidth}}
Layer & Quantity & Reference & Agreement \\
\hline
Kerr dynamics &
constants, frequencies, and separatrix &
analytic Kerr and Schwarzschild results
~\cite{Schmidt:2002qk,Bardeen:1972fi} &
$10^{-12}$--$10^{-5}$ \\

First-order fluxes &
circular, eccentric, and generic Kerr orbits &
BHPT Toolkit~\cite{BHPToolkit}, published fluxes~\cite{vandeMeent:2016pee},
and \textcolor{linkcolor}{\texttt{pybhpt}}~\cite{pybhpt} &
$10^{-5}$--$10^{-8}$ \\

Conservative sector &
circular, eccentric, and inclined redshift &
published self-force data
~\cite{WaSABI,vandeMeent:2015lxa,Barack:2011ed,Nasipak:2025tby} &
$2\times10^{-4}$--$8\times10^{-4}$ \\

Secondary spin &
linear- and quadratic-in-spin flux corrections &
Refs.~\cite{Akcay:2019bvk,Skoupy:2023lih,Rahman:2026qho} &
$4\times10^{-5}$--$4\times10^{-4}$ \\

Time-domain solver &
operator residual, quasinormal modes, sourced waveforms, and spin sources &
frequency-domain Teukolsky results and analytic QNMs
~\cite{Berti:2005ys,Chandrasekhar:1975zza,Detweiler:1980gk} &
$5\times10^{-13}$ residual; QNMs $\lesssim4\times10^{-3}$ \\

Inspiral dynamics &
0PA and 1PA Schwarzschild inspirals &
WaSABI~\cite{WaSABI} &
$\Delta\phi_{22}\propto\nu,\ \nu^2$ \\

Full waveform &
adiabatic inspiral and complete coalescence &
uncalibrated \textcolor{linkcolor}{\texttt{BHPTNRSurrogate}}
~\cite{Islam:2022laz} &
inspiral mismatch $\sim10^{-4}$ \\
\end{tabular}
\end{ruledtabular}
\end{table*}

\begin{figure}[t]
    \centering
    \includegraphics[width=\columnwidth]{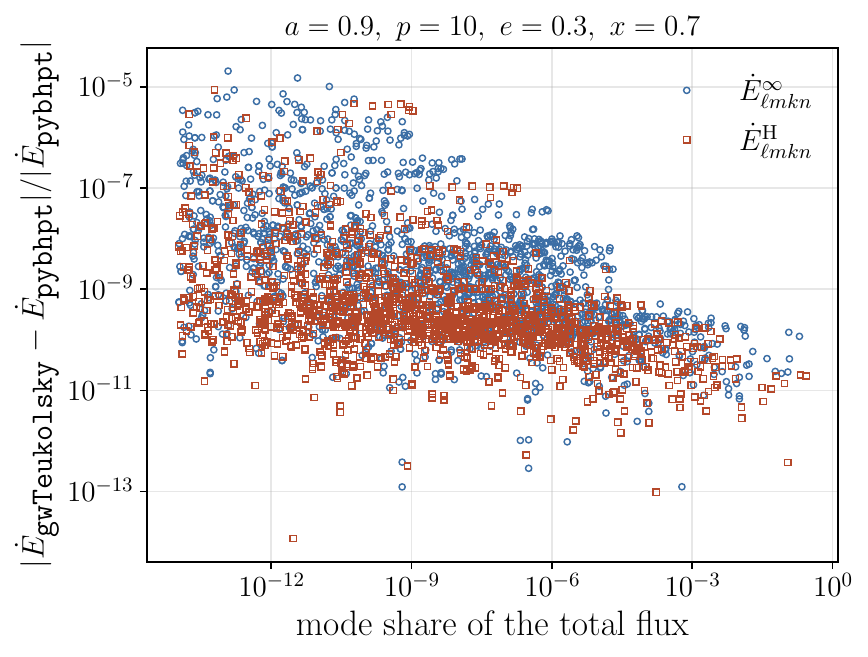}
    \caption{\label{fig:pybhpt}
    Mode-by-mode comparison of the energy flux from
    \textcolor{linkcolor}{\texttt{gwTeukolsky}} with the independent
    frequency-domain code \textcolor{linkcolor}{\texttt{pybhpt}} for the
    generic Kerr orbit $(a,p,e,x)=(0.9,10,0.3,0.7)$. Shown are the
    relative differences for $2265$ $(\ell,m,k,n)$ modes at infinity
    (circles) and at the horizon (squares). Modes carrying more than
    $10^{-6}$ of the total flux agree at the $\sim10^{-8}$ level or
    better, while the summed flux agrees to $5\times10^{-11}$. More details are provided in Appendix~\ref{app:validation}.}
\end{figure}

We separately include the leading comparable-mass dependence of the
individual waveform modes~\cite{Kidder:2007rt,Damour:2008gu}. Writing $X_i=m_i/M$, the Newtonian mass and current multipoles give
\begin{align}
c_{\ell m}(\nu)
&=
\begin{cases}
\left|X_2^{\ell-1}+(-1)^\ell X_1^{\ell-1}\right|,
& \ell+m\ {\rm even},\\
\left|X_2^\ell-(-1)^\ell X_1^\ell\right|,
& \ell+m\ {\rm odd},
\end{cases}
\nonumber\\
h_{\ell m}&\rightarrow c_{\ell m}h_{\ell m},
\qquad
\mathcal F_{\ell m}\rightarrow
c_{\ell m}^{\,2}\mathcal F_{\ell m}.
\label{eq:newtonian_prefactors}
\end{align}
These factors approach unity in the perturbative limit but become
significant already at moderate mass ratio. At $q=10$, for example, they
reduce the $(3,3)$ flux by about one third and the $(4,4)$ flux by $43\%$,
and reproduce the corresponding NR mode-flux suppression in SXS:BBH:1107
at the percent level. The
\textcolor{linkcolor}{\texttt{0PA-N}} tier applies these factors mode by
mode. This is a Newtonian finite-mass-ratio resummation rather than a
self-force correction, since higher-PN mass-ratio dependence within each
mode is not included.

\section{External validation}
\label{app:validation}

We validate \textcolor{linkcolor}{\texttt{gwKerrDynamics}} and
\textcolor{linkcolor}{\texttt{gwTeukolsky}} at several levels, from
geodesic quantities and fluxes to complete inspiral--merger--ringdown
waveforms.

The validation is organized to test the calculation at several distinct
levels. Wherever possible, we compare individual ingredients against
independent implementations or published results: Kerr orbital quantities
against analytic limits, frequency-domain fluxes against independent
Teukolsky calculations, and finite-mass-ratio corrections against published
self-force results. We then test the accumulated orbital evolution against
an independent inspiral calculation and, finally, the assembled waveform
pipeline against an existing BHPT surrogate. This separation is useful
because agreement of a complete waveform can otherwise hide compensating
errors between the dynamics and radiation. We therefore exclude comparisons
between routines within our own software suite from the external validation
summarized in Table~\ref{tab:validation}.

\begin{figure}[t]
    \centering
    \includegraphics[width=\columnwidth]{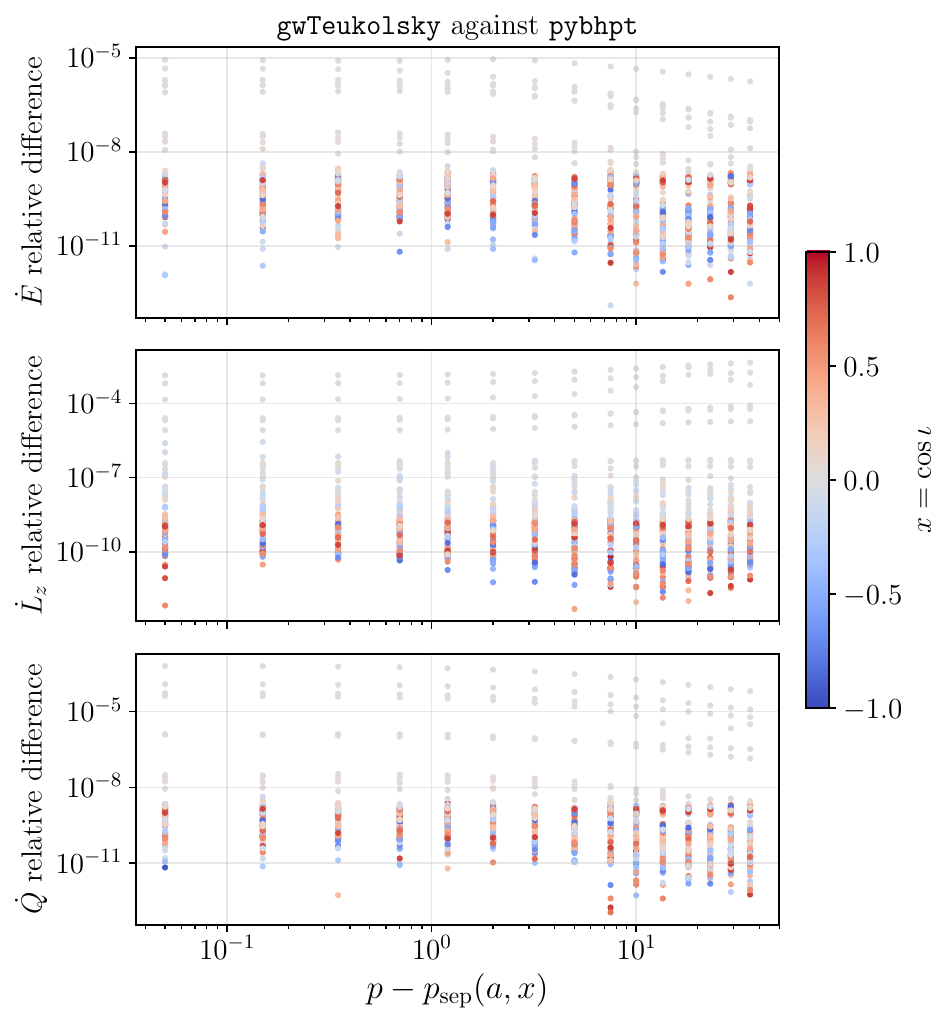}
    \caption{\label{fig:pybhpt_tables}
    Comparison of the inclined-orbit flux computations from
    \textcolor{linkcolor}{\texttt{gwTeukolsky}} with the independent
    frequency-domain code \textcolor{linkcolor}{\texttt{pybhpt}} at all
    $1605$ production points. Relative differences in $\dot E$, $\dot L_z$,
    and $\dot Q$ are shown against the distance $p-p_{\rm sep}(a,x)$ from
    the separatrix, with color indicating the inclination $x=\cos\iota$. 
    More details are provided in Appendix~\ref{app:validation}.}
\end{figure}

A direct test of the first-order flux calculation comes from
\textcolor{linkcolor}{\texttt{pybhpt}}~\cite{pybhpt}, an independent
frequency-domain Teukolsky code. For the generic Kerr orbit
$(a,p,e,x)=(0.9,10,0.3,0.7)$, we compare $2265$ individual
$(\ell,m,k,n)$ energy-flux modes at infinity and the horizon
(Fig.~\ref{fig:pybhpt}). Modes contributing more than $10^{-6}$ of the
total flux agree to $1.1\times10^{-8}$ at infinity and
$3.7\times10^{-9}$ at the horizon, while the summed flux agrees to
$5\times10^{-11}$. We repeat this comparison at all $1605$ nodes of the
inclined production tables, spanning $14$ spins from $a=0.09$ to $0.95$,
inclinations across $-1<x<1$, and
$0.05\leq p-p_{\rm sep}\leq36$ (Fig.~\ref{fig:pybhpt_tables}). Away from
the poles, $|x|>0.05$, the tables agree to $1.2\times10^{-8}$ in
$\dot E$, $2.8\times10^{-6}$ in $\dot L_z$, and $10^{-8}$ in $\dot Q$,
with median differences of a few times $10^{-10}$. The largest relative
differences occur at the $135$ nearly polar nodes, $|x|<0.05$, reaching
$9\times10^{-6}$ in $\dot E$, $4.5\times10^{-3}$ in $\dot L_z$, and
$6\times10^{-4}$ in $\dot Q$. There $\dot L_z$ approaches zero, so small
absolute differences produce large relative ones. This tests the flux
calculation for generic bound Kerr motion rather than only the circular
or equatorial limits.

\begin{figure*}[t]
    \centering
    \includegraphics[width=\textwidth]{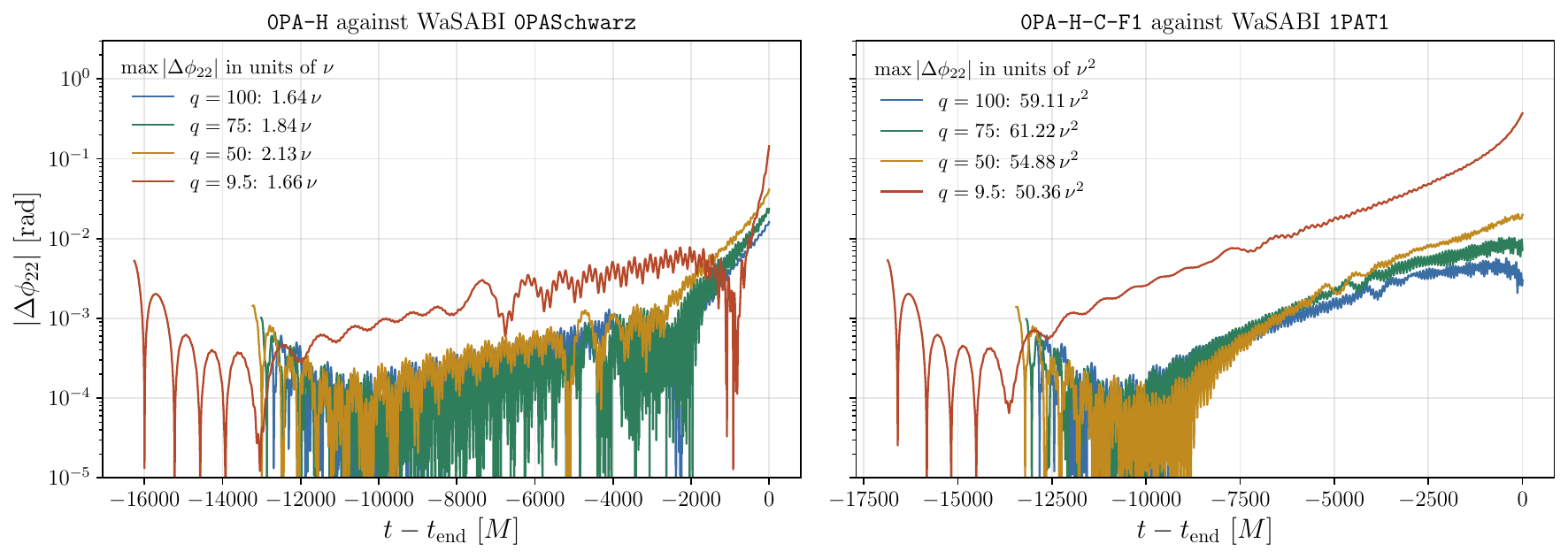}
    \caption{\label{fig:wasabi}
    Comparison with WaSABI for nonspinning Schwarzschild inspirals at
    $q=9.5$, $50$, $75$, and $100$. Each pair starts at the same radius
    and is aligned with a single time and phase shift over the first
    $3000M$. The left panel shows the phase difference using 0PA dynamics,
    for which the maximum residual scales approximately as $O(\nu)$; the
    right panel shows the 1PA comparison, where it scales as $O(\nu^2)$.
    The legend reports the maximum phase difference normalized by $\nu$
    and $\nu^2$ in the two panels, respectively. The comparison ends at
    $M\omega_{22}=0.10$, before the transition region. More details are provided in Appendix~\ref{app:validation}.}
\end{figure*}

\begin{figure*}[t]
    \centering
    \includegraphics[width=\textwidth]{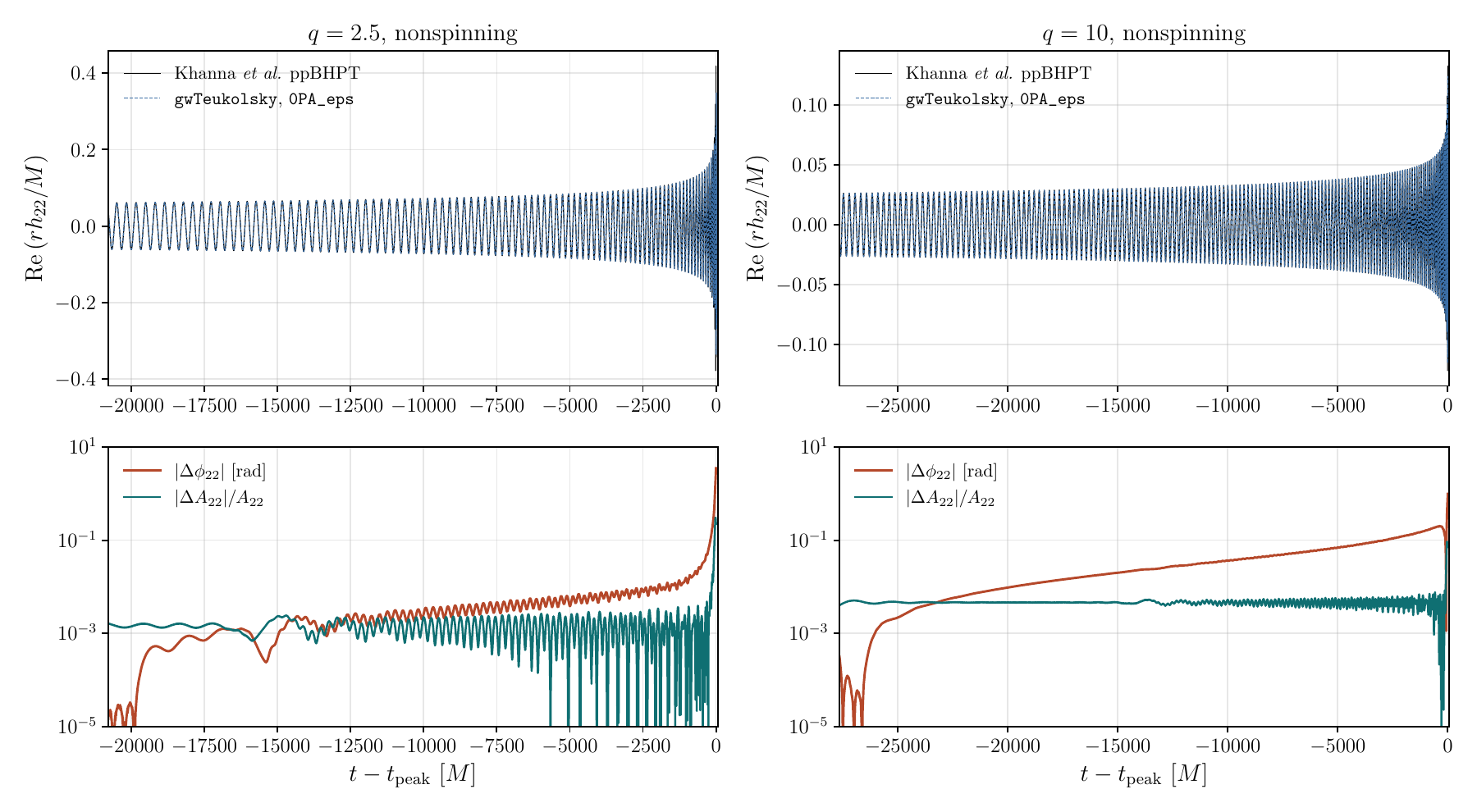}
    \caption{\label{fig:khanna}
    Direct waveform comparison between
    \textcolor{linkcolor}{\texttt{gwTeukolsky}} and the independent
    point-particle BHPT code developed by Khanna and collaborators
    ~\cite{Khanna:2004,Sundararajan:2007jg,Sundararajan:2008zm,
    Sundararajan:2010sr}, for nonspinning binaries with $q=2.5$ (left)
    and $q=10$ (right). The upper panels show the real part of the $(2,2)$
    strain; the lower panels show the absolute phase difference and
    relative amplitude difference. A single time and phase alignment is
    performed early in the inspiral, with no subsequent alignment.
    During inspiral, the amplitude agrees within $\lesssim7\times10^{-3}$; the larger
    late-time phase difference is associated primarily with the different
    transition-to-plunge prescriptions. 
    More details are provided in Appendix~\ref{app:validation}.}
\end{figure*}

Every waveform in this work is generated with the time-domain Teukolsky
solver, so we validate the field evolution separately from the
frequency-domain fluxes that drive the orbit. Substituting a validated
frequency-domain mode into the discretized Teukolsky operator gives a
residual of $5\times10^{-13}$. Starting instead from generic initial data,
the evolved field recovers the quasinormal-mode frequencies of
Ref.~\cite{Berti:2005ys} to $5\times10^{-4}$ for $a=0$ and
$4\times10^{-3}$ for $a=0.9$. At the end of a full coalescence, the
$(2,2)$ and $(3,3)$ ringdowns agree with the corresponding
Chandrasekhar--Detweiler values
~\cite{Chandrasekhar:1975zza,Detweiler:1980gk} to
$1.5\times10^{-3}$ in frequency and $1.8\times10^{-3}$ in damping rate.
For a particle on eccentric and inclined Kerr geodesics, the time-domain
field also reproduces frequency-domain snapshot waveforms, with an
amplitude ratio of $1.000$ and an overlap of $1.0000$ for each of four
modes. With a spinning secondary, the linear response of the $(2,2)$
amplitude to the pole--dipole and quadrupole sources agrees with the
frequency-domain calculation to $6\times10^{-3}$ and
$5\times10^{-4}$, respectively. 

We next compare the inspiral dynamics with WaSABI~\cite{WaSABI} for
nonspinning binaries with $q=9.5$, $50$, $75$, and $100$. We start the two
calculations at the same radius, evolve to $M\omega_{22}=0.10$ before the
transition region, and align each pair with a single time and phase shift
over the first $3000M$. Figure~\ref{fig:wasabi} shows that at 0PA the
maximum phase difference scales as
$\Delta\phi_{22}\simeq(1.6$--$2.1)\nu$, whereas including the 1PA dynamics
changes the residual to $(50$--$61)\nu^2$. The relative amplitude
difference remains below $4\times10^{-3}$ for $q\geq50$ and reaches
$1.7\times10^{-2}$ at $q=9.5$. The 1PA comparison is not an independent
validation of the underlying self-force data, since the second-order
dissipative and conservative inputs used here are derived from WaSABI data.
It instead tests their conversion into our evolution variables and their
subsequent propagation through the inspiral. The small $O(\nu)$ difference
at 0PA is not traced to a unique source; possible contributions include the
time-domain evolution of the sourced field and post-leading terms in our
composite worldline. The change from an $O(\nu)$ residual at 0PA to
$O(\nu^2)$ at 1PA is nevertheless consistent with the expected difference
beyond the perturbative order shared by the two calculations.

Finally, we test the complete waveform pipeline against the point-particle
BHPT waveforms produced in the code developed by Khanna and
collaborators~\cite{Khanna:2004,Sundararajan:2007jg,Sundararajan:2008zm,Sundararajan:2010sr},
as used in Refs.~\cite{Rifat:2019ltp,Islam:2022laz,Rink:2024swg}.
Figure~\ref{fig:khanna} compares our two longest nonspinning
\textcolor{linkcolor}{\texttt{0PA\_eps}} waveforms, at $q=2.5$ and $q=10$,
with theirs at the same mass ratios, in total-mass units. After discarding
the first $1500M$ of the common span, where both records still carry their
initial transients, we fit a single time and phase shift over the next
$1000M$, with no subsequent alignment. The comparison then covers about
$20{,}800M$ ($\sim96$ GW cycles) at $q=2.5$ and $27{,}600M$
($\sim155$ cycles) at $q=10$, through the transition, plunge, and
ringdown. The inspiral amplitudes agree within $4\times10^{-3}$ and
$8\times10^{-3}$. The $(2,2)$ phase difference stays below $0.03$ rad
until $500M$ before the peak at $q=2.5$, and grows to $0.19$ rad at
$q=10$. The larger differences in the last $\sim200M$ arise from the
distinct transition-to-plunge prescriptions used in the two calculations.
This comparison tests the assembled pipeline (orbital evolution,
transition, time-domain radiation, and waveform construction) rather than
any one perturbative ingredient in isolation.

We then extend this test to multiple binaries with a spinning primary using our
\textcolor{linkcolor}{\texttt{0PA\_eps}} waveforms. To facilitate the
comparison, we use the uncalibrated
\textcolor{linkcolor}{\texttt{BHPTNRSur2dq1e3}}
surrogate, built from waveforms produced by the pipeline developed by
Khanna and collaborators. We compare the $(2,2)$ mode over the adiabatic
inspiral, from the beginning of each waveform to the start of our
transition. For the $33$ systems within the surrogate's training region,
$3\leq q\leq1000$ and $|\chi_1|\leq0.8$, our sample spans
$3\leq q\leq32$ and $-0.8\leq\chi_1\leq0.8$, with comparison windows of
$6100M$--$12{,}400M$, or $41$--$108$ GW cycles. With one time and phase
shift fitted over the full comparison window, we find a median time-domain
$L_2$-norm error of $9.7\times10^{-5}$, ranging from
$4.5\times10^{-5}$ to $4.0\times10^{-4}$. If instead we align only over
the first $1000M$, the accumulated $(2,2)$ phase difference has a median
of $0.20$ rad at the end of the inspiral and remains below $0.42$ rad.
The median relative amplitude difference is $4.4\times10^{-3}$, with a
maximum of $2.6\times10^{-2}$ across the sample.
For comparison, $19$ nonspinning
\textcolor{linkcolor}{\texttt{0PA\_eps}} waveforms with
$2.5\leq q\leq20$ are tested against
\textcolor{linkcolor}{\texttt{BHPTNRSur1dq1e4}} over $32$--$97$ GW
cycles. We find a median $L_2$-norm error of $1.1\times10^{-4}$, with
none above $1.7\times10^{-4}$, a median accumulated phase difference of
$0.21$ rad, and a median relative amplitude difference of
$3\times10^{-3}$.
Our \textcolor{linkcolor}{\texttt{0PA\_eps}} waveforms calculation and the code developed by Khann and
collaborators use the same adiabatic
point-particle physics but are independent implementations, with different
flux computations, source discretizations, initial data, and transition
prescriptions. The comparison also inherits the modeling error of the
surrogates themselves: at $\chi_1=0$, the two uncalibrated surrogates
differ in amplitude by $2$--$6\times10^{-3}$, comparable to the
differences found here.

\begin{table*}[t]
\caption{\label{tab:agentic}
Examples of errors that survived internal tests and the independent checks
that exposed them. More details are provided in Appendix~\ref{app:agentic}.}
\footnotesize
\begin{ruledtabular}
\begin{tabular}{p{0.31\textwidth}p{0.34\textwidth}p{0.30\textwidth}}
Problem & Independent check & Validation principle \\
\hline

A transition diagnostic normalized by $E$ rather than
$E-E_{\rm ISCO}$ turned a mismatch of $0.85$ into
$3.4\times10^{-3}$. &
Comparison with
\textcolor{linkcolor}{\texttt{BHPTNRSurrogate}}
~\cite{Islam:2022laz} and its mass normalization. &
Use an independent calculation or analytic limit when available. \\

The generic conservative correction agreed with the equatorial result but
was wrong by factors of $1.8$--$4.7$ for inclined orbits. &
Published inclined-orbit redshifts~\cite{Nasipak:2025tby} and Schwarzschild
rotational invariance. &
Do not validate a term only where its nontrivial dependence disappears. \\

For some MAYA binaries, nominal rather than realized mass ratios produced
$1$--$2$ rad of accumulated phase difference. &
Audit of $163$ NR records against their metadata. &
Take physical parameters directly from the simulation record. \\

\end{tabular}
\end{ruledtabular}
\end{table*}

\section{Numerical resolution and convergence}
\label{app:resolution}

\begin{figure}[t]
    \centering
    \includegraphics[width=\columnwidth]{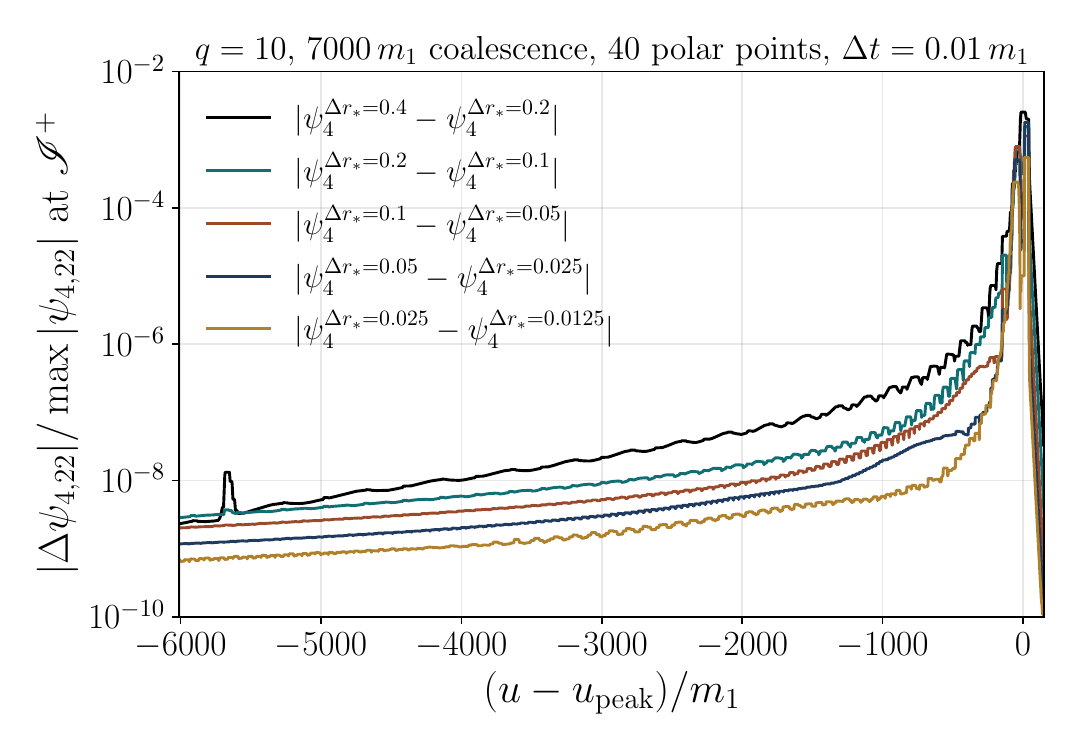}
    \caption{\label{fig:convergence_long}
    Radial self-convergence of $\psi_4^{22}$ for a nonspinning
    $q=10$ coalescence, evolved with $40$ polar points and
    $\Delta t=0.01\,m_1$. Each curve shows the difference between
    successive radial resolutions, normalized by the peak amplitude,
    after removing the leading finite-radius correction and
    interpolating onto a common retarded-time grid. A running maximum
    over $30\,m_1$ is shown. The differences decrease with resolution
    during inspiral but approach a floor near merger.
    Further details are given in Appendix~\ref{app:resolution}.}
\end{figure}

We examine numerical resolution for two $q=10$ coalescences. For the
nonspinning case, we evolve the same
\textcolor{linkcolor}{\texttt{0PA\_eps}} worldline from
$r_0=14.7\,m_1$ through transition and plunge, covering about
$7000\,m_1$ and $28$ orbits. Keeping the worldline fixed lets us
measure numerical differences in the radiative evolution alone.
We use $40$ polar points, $\Delta t=0.01\,m_1$, and five radial
spacings from $\Delta r_*=0.4$ to $0.025\,m_1$.
Figure~\ref{fig:convergence_long} compares $\psi_4^{22}$ at successive
resolutions. Before taking the differences, we remove the leading
finite-radius correction from the waveforms and interpolate them
onto a common retarded-time grid.

The differences decrease with radial resolution throughout the
inspiral. Early on, they are only $10^{-9}$--$10^{-8}$ of the peak
amplitude; up to $300\,m_1$ before merger, the relative $L_2$
difference between the two finest resolutions is
$8.6\times10^{-7}$. The observed convergence is slower than the
formal fifth-order radial stencil, so other numerical effects
appear to limit the accuracy. Near merger, the differences grow
and approach a floor of about $5\times10^{-4}$, associated with
the transition-to-plunge handover and the end of the worldline.
We also halve the time step at $\Delta r_*=0.05\,m_1$, from
$0.01$ to $0.005\,m_1$. This changes $\psi_4^{22}$ by less than
$4.1\times10^{-9}$ during inspiral, two to three orders of
magnitude below the radial differences. Near merger, the
time-step difference reaches $1.6\times10^{-4}$, showing that
time integration also contributes to the late-time residual.

We repeat the resolution test for an eccentric, inclined coalescence
with $q=10$, $a=0.7$, $e_0=0.2$, and $x_0=0.87$, evolved through
transition and plunge. We compare radial spacings
$\Delta r_*=0.1$, $0.05$, and $0.025\,m_1$ with $20$ polar points,
and repeat the first two with $32$ points. The strain waveforms
overlap at all five resolutions. Relative to the finest run, the
peak difference in $\psi_4^{22}$ decreases from
$9.0\times10^{-4}$ at $\Delta r_*=0.1\,m_1$ to
$3.3\times10^{-4}$ at $0.05\,m_1$, with the largest differences
near merger. During inspiral, they remain at the
$10^{-6}$--$10^{-5}$ level, and increasing the polar resolution
makes little difference.

\section{Perturbative ladder at larger mass ratio}
\label{app:ladder-q15}

\begin{figure}[t]
    \centering
    \includegraphics[width=\columnwidth]{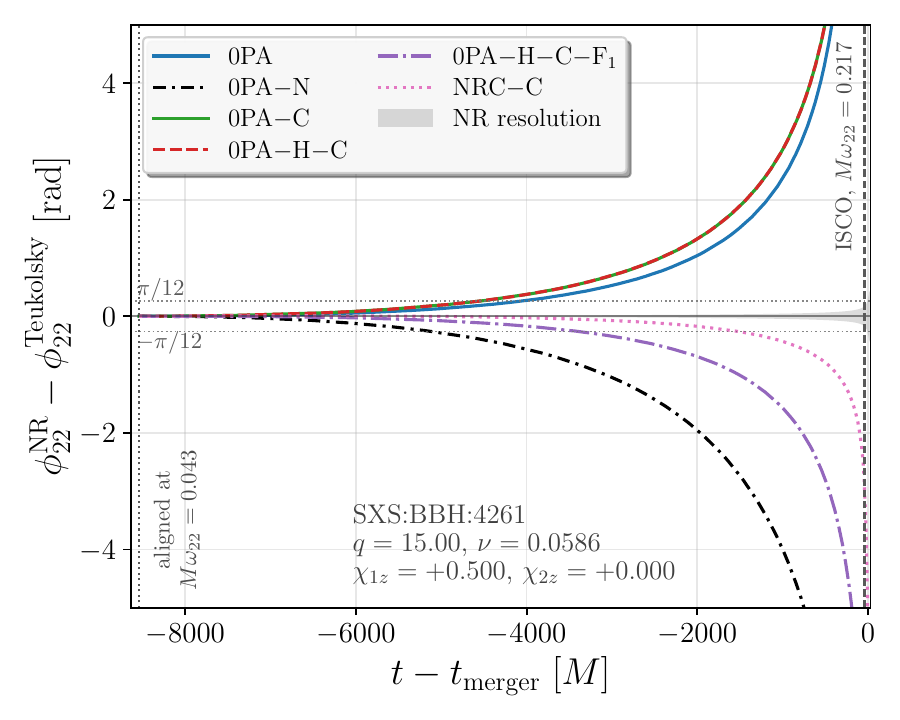}
    \caption{\label{fig:ladder-q15}
    Perturbative ladder for the aligned-spin binary SXS:BBH:4261,
    with $q=15$, $\nu=0.0586$, $\chi_{1z}=0.5$, and $\chi_{2z}=0$.
    We show the $(2,2)$ phase difference between NR and the time-domain
    Teukolsky waveforms after alignment at $M\omega_{22}=0.043$.
    Starting from \textcolor{linkcolor}{\texttt{0PA}}, the curves show
    the effect of the Newtonian mass-asymmetry correction
    \textcolor{linkcolor}{\texttt{0PA-N}}, the conservative correction
    \textcolor{linkcolor}{\texttt{C}}, horizon absorption
    \textcolor{linkcolor}{\texttt{H}}, and the second-order dissipative
    correction \textcolor{linkcolor}{\texttt{F1}}, together with the
    NR-calibrated \textcolor{linkcolor}{\texttt{NRC-C}} prescription.
    The gray band denotes the NR resolution uncertainty, the horizontal
    dotted lines mark $|\Delta\phi_{22}|=\pi/12$, and the vertical dotted
    line marks the Kerr ISCO. More details are provided in 
    Appendix~\ref{app:ladder-q15}.}
\end{figure}

Figure~\ref{fig:ladder-q15} gives a second example of how the individual
finite-mass-ratio corrections modify the waveform, now for
SXS:BBH:4261, with $q=15$, $\chi_{1z}=0.5$, and $\chi_{2z}=0$.
The waveforms are aligned at $M\omega_{22}=0.043$, roughly $8000M$
before merger. The qualitative behavior differs from the equal-mass
example in Fig.~\ref{fig:ladder}. The bare
\textcolor{linkcolor}{\texttt{0PA}} evolution accumulates a positive
phase difference, while the Newtonian mass-asymmetry correction
\textcolor{linkcolor}{\texttt{0PA-N}} overcorrects it and produces a
substantially larger dephasing with the opposite sign. By contrast, the
conservative correction moves the perturbative waveform closer to NR,
with horizon absorption producing only a small additional change. Adding
the second-order dissipative correction
\textcolor{linkcolor}{\texttt{F1}} further reduces the phase difference
over most of the inspiral. The NR-calibrated
\textcolor{linkcolor}{\texttt{NRC-C}} prescription gives the closest
agreement until late in the evolution. This example illustrates that the
different finite-mass-ratio prescriptions are not interchangeable:
corrections that incorporate comparable-mass dependence at the waveform
or flux level can move the result in substantially different directions,
while the conservative and dissipative self-force corrections act directly
on the accumulated orbital dynamics.

Normalizing the accumulated phase difference by the number of NR orbits
since alignment makes the mass-ratio trend particularly clear. At $1000M$
before merger, the \textcolor{linkcolor}{\texttt{0PA}},
\textcolor{linkcolor}{\texttt{0PA-H-C}}, and
\textcolor{linkcolor}{\texttt{NRC}} waveforms dephase per orbit by factors
of $3.1$--$3.8$ less at $q=15$ than at $q=1$, close to the factor $4.3$
decrease in $\nu$ expected for an approximately $O(\nu)$ error. The main
exception is \textcolor{linkcolor}{\texttt{F1}}, which slightly
overcorrects at $q=15$, consistent with applying the available
Schwarzschild second-order flux to a prograde primary with
$\chi_{1z}=0.5$.

\section{Agentic development of
\textcolor{linkcolor}{\texttt{gwKerrDynamics}} and
\textcolor{linkcolor}{\texttt{gwTeukolsky}}}
\label{app:agentic}

The two software packages used in this work were developed with extensive
use of AI coding agents. Over the 35 days following the initial development plan,
\textcolor{linkcolor}{\texttt{gwKerrDynamics}} and
\textcolor{linkcolor}{\texttt{gwTeukolsky}} grew to approximately
$3.4\times10^4$ lines across $120$ modules, with $623$ tests, and were used
to generate the $1525$ waveforms in the \vc{}. The agents were used for
implementation, test generation, code execution, bookkeeping, and
exploratory debugging. We specified the physical formulation, conventions,
validation criteria, and interpretation. Because agents could readily
produce plausible explanations for discrepancies, we treated a diagnosis
as provisional until a calculation could distinguish it from competing
explanations.

Development began with the physics rather than the software architecture.
Through an extended dialogue with GPT-5.6 Sol, we assembled a $28$-page
technical guide, \textcolor{linkcolor}{\texttt{GUIDE.pdf}}, covering the
formulation, numerical methods, conventions, and validation strategy for
the time-domain Teukolsky calculation
~\cite{Khanna:2004,Sundararajan:2007jg,Sundararajan:2008zm,
Sundararajan:2010sr}. Claude Opus 4.8 was then used to translate this into
the $684$-line \textcolor{linkcolor}{\texttt{PLAN.md}}, which divided the
calculation between \textcolor{linkcolor}{\texttt{gwKerrDynamics}} and
\textcolor{linkcolor}{\texttt{gwTeukolsky}} and specified implementation
milestones and validation targets. Development proceeded from vacuum
Teukolsky evolution to sourced evolution, inspiral, transition, and plunge,
with finite-mass-ratio ingredients added individually.

\begin{figure}[t]
    \centering
    \includegraphics[width=\columnwidth]{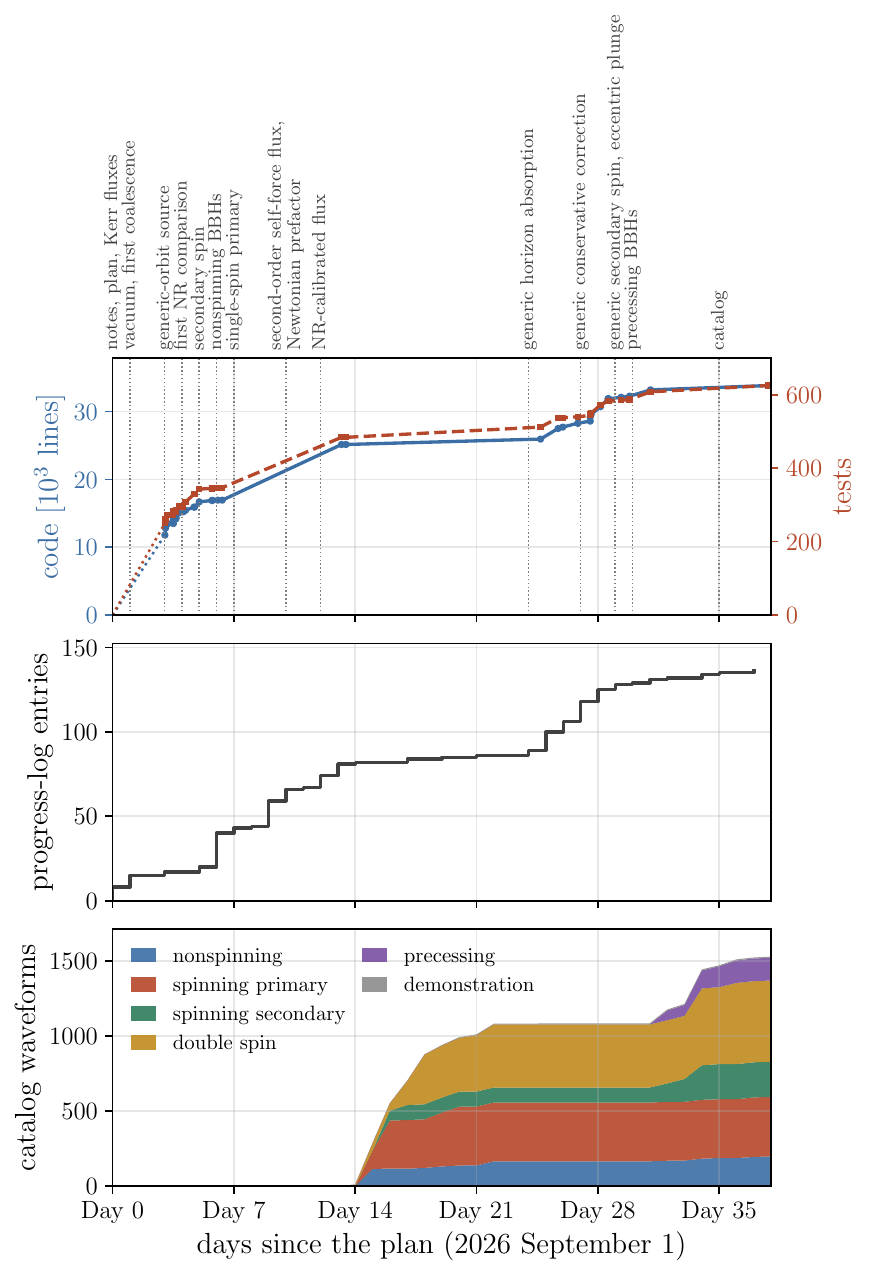}
    \caption{\label{fig:agentic-progress}
    Five weeks of agent-assisted development beginning September 1, 2026.
    Top: source-code size and test count, with major implementation
    milestones marked vertically. Middle: cumulative entries in the
    \textcolor{linkcolor}{\texttt{PROGRESS.log}} files. Bottom: cumulative
    catalog waveforms by binary family. Development, validation, and
    production proceeded concurrently. More details are provided in Appendix~\ref{app:agentic}.}
\end{figure}

We maintained a persistent development record rather than relying on the
agents' conversational context. The two
\textcolor{linkcolor}{\texttt{PROGRESS.log}} files contain $10{,}212$
lines in $135$ dated entries recording implementations, failures, diagnoses,
and quantitative tests, while the package
\textcolor{linkcolor}{\texttt{README}} files collect external validation
results. Figure~\ref{fig:agentic-progress} shows the growth of the code,
tests, development record, and waveform catalog. Implementation, validation,
and production were interleaved: new physical ingredients were tested
before being propagated through the catalog.

The surviving Claude Code session logs provide a complementary measure of
the scale of agent use. From Day 4 onward they contain $14{,}532$ model
calls across $13$ sessions, comprising approximately $7.25$ billion
processed tokens. About $98\%$ of this total consists of cached context
reread across successive calls; excluding these cache reads gives roughly
$126$ million tokens written to the cache and $16$ million output tokens.
These numbers are a lower bound on the complete development history because
the sessions from Days 0--3 are no longer available and the earlier GPT-5.6
conversations used to construct
\textcolor{linkcolor}{\texttt{GUIDE.pdf}} are not included. They are also
not a measure of code development alone: the same sessions were used for
production runs, validation, data analysis, figures, and preparation of
this manuscript.

The central difficulty was not generating code but deciding when a
calculation was scientifically trustworthy. Unit and regression tests can
preserve an incorrect implementation as effectively as a correct one. We
therefore emphasized checks independent of the implementation: published
results, independent codes, analytic limits, symmetries, and exact
properties of the underlying problem. Several errors survived the internal
test suite while producing physically plausible outputs. Three examples are
given in Table~\ref{tab:agentic}.
The inclined conservative correction illustrates this strategy. Agreement
in the equatorial limit initially suggested that the implementation was
correct, but published inclined-orbit data showed errors by factors of
$1.8$--$4.7$. A simpler test resolved the ambiguity: in Schwarzschild, an
inclined circular orbit is related to an equatorial orbit by a rotation, so
scalar observables cannot depend on inclination. The code failed this test,
ultimately revealing three separate defects. We subsequently looked for a
symmetry, limiting case, independent implementation, or other pass/fail
calculation before asking an agent to interpret the detailed structure of a
residual.

This experience suggests a practical division of labor for agentic
scientific software. Agents substantially reduced the cost of implementing
alternative formulations, writing tests, running calculations, inspecting
outputs, and exploring possible explanations for failures. They were less
useful for deciding whether a numerically plausible result was physically
correct. Reliability instead came from coupling this speed to validation
criteria independent of the generated implementation: analytic limits,
physical symmetries, published data, and independent codes. Our workflow
was therefore not ``generate and trust,'' but generate, test, challenge, and only then use for physics.

\bibliography{references}

\end{document}